\documentclass[pdflatex,sn-mathphys-num]{sn-jnl}
\usepackage{graphicx}%
\usepackage{multirow}%
\usepackage{amsmath,amssymb,amsfonts}%
\usepackage{amsthm}%
\usepackage{mathrsfs}%
\usepackage[title]{appendix}%
\usepackage{xcolor}%
\usepackage{textcomp}%
\usepackage{manyfoot}%
\usepackage{booktabs}%
\usepackage{algorithm}%
\usepackage{algorithmicx}%
\usepackage{algpseudocode}%
\usepackage{listings}%

\usepackage{nicefrac}

\usepackage{comment}
\usepackage{tikz}

\theoremstyle{thmstyleone}%

\theoremstyle{thmstyletwo}%

\theoremstyle{thmstylethree}%
\newcommand{\h}{\mathsf{h}}
\newcommand{\ep}{\mathbf{e}_{\varphi}}
\newcommand{\et}{\mathbf{e}_{\theta}}
\newcommand{\er}{\mathbf{e}_r}
\newcommand{\s}{\mathbb{S}}

\newcommand{\bu}{\boldsymbol{u}}

\newcommand{\ea}{\mathbf{e}_{1}}
\newcommand{\eb}{\mathbf{e}_{2}}
\newcommand{\ec}{\mathbf{e}_{3}}

\newcommand{\td}{\theta^{\dag}}
\newcommand{\pd}{\varphi^{\dag}}

\newcommand*\Laplace{\mathop{}\!\mathbin\bigtriangleup}

\newcommand\D{\mathcal{D}}

\begin{document}

\title[Article Title]{On the modeling of nonlinear wind-induced ice-drift ocean currents at the North Pole}


\author{\fnm{Christian} \sur{Puntini}}\email{christian.puntini@univie.ac.at}

\affil{\orgdiv{Faculty of Mathematics}, \orgname{University of Vienna}, \orgaddress{\street{Oskar-Morgenstern-Platz 1}, \city{Vienna}, \postcode{1090},  \country{Austria}}}

\abstract{Starting from the governing equations for geophysical flows, by means of a thin-shell approximation and a tangent plane approximation, we derive the equations describing, at leading order, the nonlinear ice-drift flow for regions centered around the North Pole. An exact solution is derived in the material/Lagrangian formalism, describing a superposition of oscillations, a mean Ekman flow and a geostrophic current.}

\keywords{Arctic Ocean, Navier-Stokes equations, Ice-drift current, Asymptotic methods, Ekman spiral, Rotating spherical coordinates}


\pacs[MSC Classification]{76M45, 76U60, 35B30, 35C05, 35Q30, 35Q35}

\maketitle

\section{Introduction}
\begin{figure}
    \centering
    \includegraphics[width=0.7\linewidth]{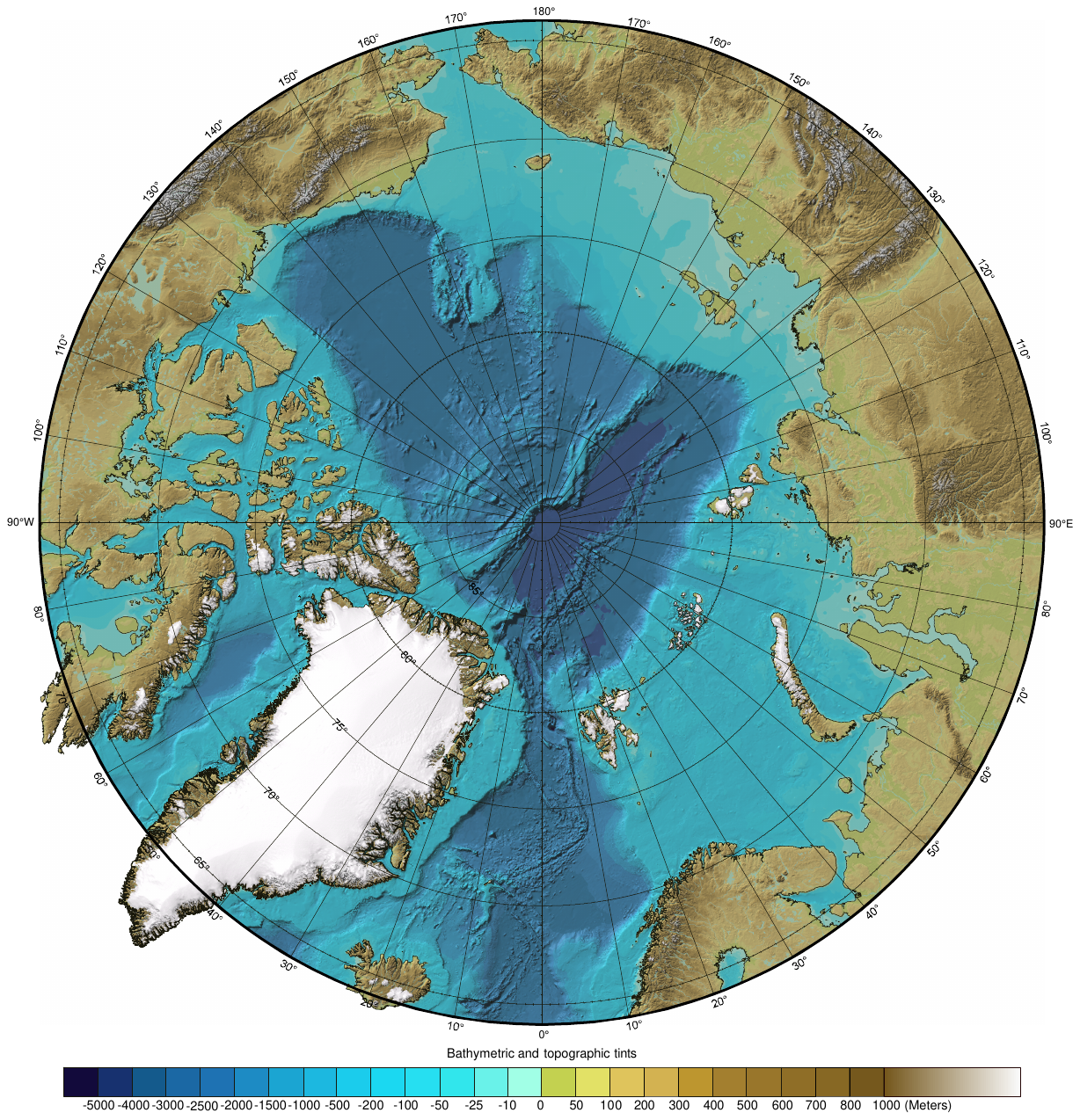}
    \caption{IBCAO bathymetry map of the Arctic Ocean, indicating circles of latitude 5 degrees apart. From \url{https://www.ngdc.noaa.gov/mgg/bathymetry/arctic/currentmap.html}. Credit \cite{IBCAO}}
    \label{fig-ARCTIC}
\end{figure}

\begin{figure}
    \centering
    \includegraphics[width=\linewidth]{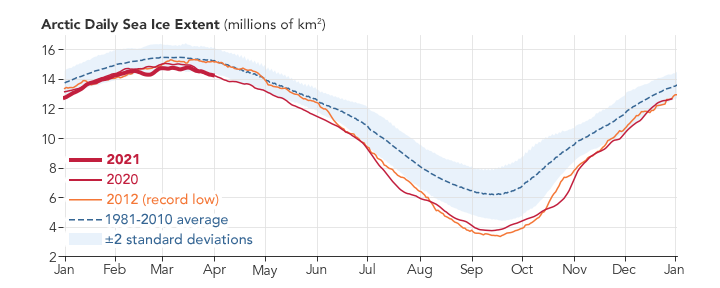}
    \caption{Arctic daily sea ice extent (millions of $km^2$), up to April 2021. From NASA’s Earth Observatory \url{https://earthobservatory.nasa.gov/world-of-change/sea-ice-arctic}.}
    \label{fig-ice}
\end{figure}

\begin{figure}
    \centering
    \includegraphics[width=\linewidth]{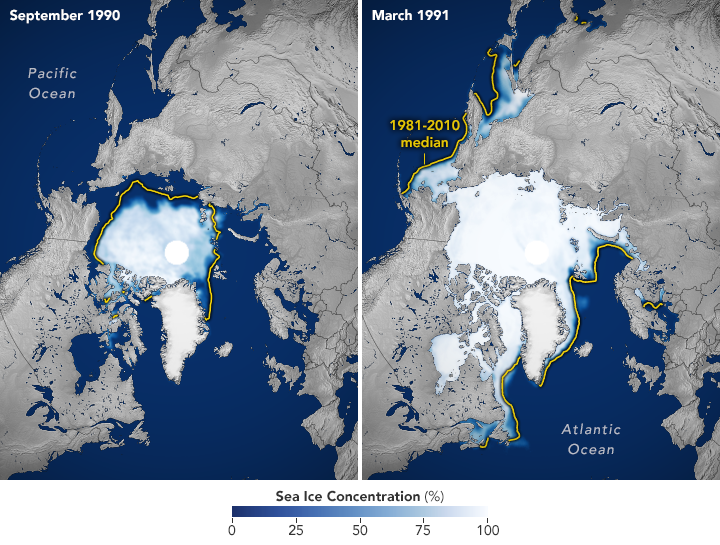}
    \caption{Average concentration of sea ice for the first available annual maxima and minima. From NASA’s Earth Observatory \url{https://earthobservatory.nasa.gov/world-of-change/sea-ice-arctic}.}
    \label{fig-ice1990}
\end{figure}

\begin{figure}
    \centering
    \includegraphics[width=\linewidth]{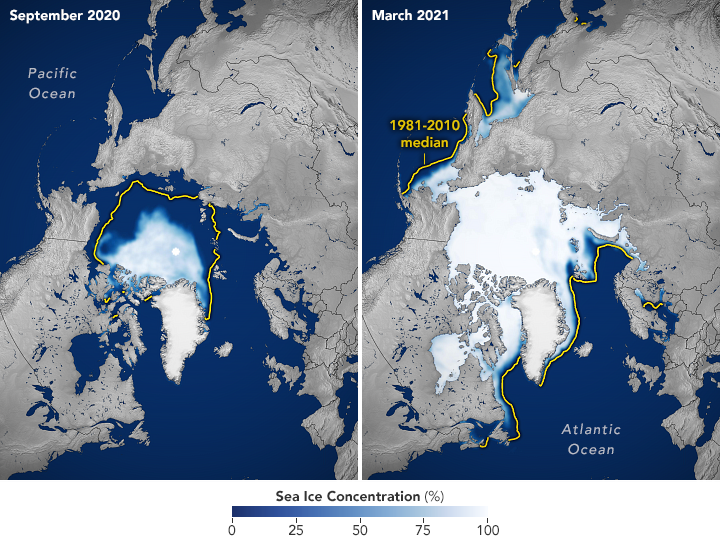}
    \caption{Average concentration of sea ice for the latest available annual maxima and minima. From NASA’s Earth Observatory \url{https://earthobservatory.nasa.gov/world-of-change/sea-ice-arctic}.}
    \label{fig-ice2021}
\end{figure}

The Arctic Ocean, centered over the North Pole, is a deep basin with maximum water depths exceeding $4000 \,m$, surrounded by extensive continental shelves with maximum depths between $300 \,m$ and $400 \,m$. This ocean basin connects to the Pacific Ocean via the shallow Bering Strait, with depths of less than $45 \,m$, and to the Atlantic Ocean through the deeper Greenland Sea (maximum depths of $2600 \,m$) and Denmark Strait (maximum depths of $650\,m$). Additionally, shallow channels (with maximum depths under $200\,m$) within the Canadian Arctic and Svalbard archipelagos provide further connections to the Atlantic. See Figure \ref{fig-ARCTIC}.\\
Atmospheric circulation plays a dominant role in driving the surface circulation of the ocean. Anticyclonic (clockwise) atmospheric flow over the Arctic Ocean basin sets in motion the surface current circulation, represented by the Beaufort Sea Gyre and the Transpolar Drift. These surface currents transport Arctic Water, characterized by temperatures ranging from \(-1.8^\circ C\) to \(3^\circ C\) and salinities from $31\%$ to $34\%$, across the Arctic Ocean basin. Most of this water is discharged into the North Atlantic Ocean via the East Greenland Current and the Baffin Current.\\
Sea ice is a dominant feature of polar seas, covering less than $10\%$ of the world’s oceans, with approximately $40\%$ of the world's sea ice located within the Arctic Ocean basin. \\
The presence of sea ice in the Arctic Ocean and its marginal seas, including Baffin Bay, Hudson Bay, and the Barent Sea has several important effects on the physical oceanography of the region: the surface water temperature in this region remains near the freezing point dictated by its salinity, while the formation of sea ice expels salt in a process known as brine rejection, which increases surface water density and drives the thermohaline circulation. Additionally, winds  transfer momentum from the atmosphere to the ocean surface through the sea ice cover, while the seasonal variable albedo of sea ice modulates the absorption and reflection of sunlight, affecting the energy exchange at the ocean surface and influencing the melting of the sea ice.\\
The northernmost part of the Arctic Ocean is entirely covered by a thin layer of sea ice, about $2\, m$  thick during the winter,  with summer sea-ice extent typically around one third of the winter extent. The winter maximum occurs in March, while the sea-ice minimum is in September (see \cite{TM2020} and Figure \ref{fig-ice}).
Since the start of the satellite record in 1979, a linear trend indicates that summer (September) sea-ice surface has been declining at a rate of about $1$ million $km^2$ per decade, with recent years seeing only about 4.5 million $km^2$ of sea ice in September. See Figures \ref{fig-ice1990}, \ref{fig-ice2021}. This loss in sea-ice area is accompanied by a decrease in sea-ice volume, with a shift to thinner, more mobile sea-ice packs. In the 1980s, average winter [fall] sea-ice thickness was around $3.6\, m$ [$2.7\,m$], while in 2018 it had decreased to about $2\,m$ [$1.5\,m$].  The amount and mobility of sea ice play a significant role in driving large-scale ocean circulation, as sea ice acts as a critical mediator of wind stress in the Arctic. In addition, changes in Arctic Ocean conditions, such as warming, freshening, and shifts in stratification, circulation dynamics, and momentum transfer to the ocean are closely linked to these sea-ice changes (see \cite{TM2020} and the references therein).
\newline
Given the evident importance of the Arctic Ocean, it is therefore crucial to understand the variety of phenomena characterizing this basin. The difficulty of physical observations (due to harsh meteorological conditions and to the presence of sea-ice) makes the theoretical approach based on the careful (asymptotic) analysis of the governing equations of fluid mechanics particularly important (and suitable) (see \cite{Johnson2022} for a survey on this method). Such an approach was recently pursued for the study of the dynamics of the Beaufort Gyre (see e.g. \cite{CJ2023}) and of the Transpolar Drift Current (see e.g. \cite{CJ2023}, \cite{CJ2024} or \cite{Johnson2024}).\\
In this article, we study, adopting this aforementioned approach, the nonlinear flows under the sea-ice cover induced by wind (which exerts a stress on the ice sheet, in turn inducing a reduced stress on the water surface), extending the work in \cite{Constantin2022} and \cite{Constantin2022NOTE}, which was limited to regions outside the Amundsen Basin (where the North Pole is located, thus avoiding the singularity issue of the convergence of the meridians at the pole linked to the use of classical spherical coordinates). To avoid this inconvenience, we will adopt the rotated spherical coordinates (having the singularity at the Equator) developed in \cite{CJ2023} and \cite{CJChapter}.\\
At latitudes away from the equator, the oceanic flow below a depth of about $100\,m$ typically features a geostrophic balance. However, near the surface, wind-driven turbulence transfers momentum into the ocean, generating currents that overlay the underlying geostrophic flow. The resulting force balance in these currents involves the Coriolis force and the frictional forces induced by the wind. This region, where these frictional effects are significant, is known as the Ekman layer, named after V. W. Ekman, who first developed a mathematical model, in 1905 \cite{Ekman1905}, to describe the behavior of wind-driven surface currents in order to explain the observations made by F. Nansen during the 1893-1896 Arctic expedition on board of the Fram vessel that sea ice drifts somewhat to the right of the prevailing wind direction (see \cite{JenkinsB2006} for a summary of Ekman's work). \\
Ekman's explicit solution is applicable to a uniform, steady wind blowing over a homogeneous ocean with a constant eddy viscosity. In this scenario:  the surface flow is oriented at a $45^{\circ}$ angle to the wind, with the direction being to the right in the Northern Hemisphere and to the left in the Southern Hemisphere, and as one dives deeper into the water, the current speed gradually decreases, and the direction of flow rotates further away from the wind, following a spiral pattern. On average, the wind-driven motion, known as Ekman transport, occurs at a right angle to the wind direction,  to the right in the Northern Hemisphere and to the left in the Southern Hemisphere.
These conclusions hold for cases with depth-dependent eddy viscosity (see \cite{Constantin2020}), though the first point only qualitatively matches observations, with deflection angles observed in the range of $10^{\circ}$ to $75^{\circ}$ (see \cite{Roberti} and the references therein). The discrepancy between the predicted and observed deflection angles can be mostly attributed to the assumption of constant vertical eddy viscosity in the original model (see \cite{Roberti} for a three-valued piecewise-constant eddy viscosity, or \cite{ShriraJFM} for depth and time dependent eddy viscosity).\\
As eddy viscosity is just a simple model to ``close" the turbulence problem, based on the hypothesis that Reynolds stresses are proportional to the mean velocity gradient and representing no physical characteristic of the fluid (but being function of flow conditions), there were numerous attempts to improve the Ekman solution by choosing a better turbulence model (see \cite{Soloviev2006}). We work with the assumption of constant eddy viscosity. The exploration of more intricate models will be the object of subsequent investigations.\\
Extending \cite{Constantin2022} and \cite{Constantin2022NOTE}, we provide an exact description of the nonlinear flow, which consists of oscillations superimposed on an Ekman mean spiraling current. However, in our case, the stress on the ocean surface is not directly generated by the wind but rather by the ice motion, whereas in \cite{Constantin2022} and \cite{Constantin2022NOTE}, the stress is induced by the wind, ice, or a combination of both. Differently from \cite{Constantin2022} and \cite{Constantin2022NOTE} our solution also features a background geostrophic flow, due to the presence of the Transpolar Drift Current.\\
\\
\noindent
The paper is structured as follows: in Section \ref{sec-govEQNS}, after recalling the set of rotated spherical coordinates developed in \cite{CJ2023} suitable for the analysis of fluid flows in regions centered around the poles (where classical spherical coordinates fail) we derive the governing equations of our interest, starting from the Navier-Stokes and continuity equations via a thin-shell approximation coupled with a tangent plane approximation. We also describe the boundary conditions associated with the presence of ice and its interaction with the water beneath it.    \\
In Section \ref{sec-Lag} we present the solution for the ice-drift nonlinear flow, obtained via a Lagrangian formulation of the flow variables. Such a solution features oscillations superimposed on an Ekman-type mean spiraling current and on the background geostrophic component of the Transpolar Drift Current. Moreover, it will be shown that the boundary conditions uniquely determine the surface
 current, therefore explaining how, given accurate field data about the ice motion, the mean surface current is connected to the ice-sea stress.\\
 Finally, we conclude with a discussion of the results in Section \ref{Sec-Discussion}.\\
In Appendix \ref{appA} we review the basics of differential geometry, with a focus on orthogonal curvilinear coordinate systems, in order to provide explicit expressions of the differential operators used in the subsequent derivation of the Navier-Stokes and continuity equations in their most general form in classical spherical coordinates (Appendix \ref{classical spherical coordinates}) and rotated spherical coordinates (Appendix \ref{rotated spherical}), suitable for the study of the geophysical fluid dynamics of oceans and of the atmosphere. Only two assumptions will be made: the fluid is Newtonian and the viscosity varies only with depth.

\section{Governing equations}\label{sec-govEQNS}
We assume the Earth to be a sphere of radius $R'\approx6371\ km$, disregarding the exact form of Earth's geoid, though appropriate adjustments to its shape can be incorporated (see \cite{CJ2021}), and define a Cartesian coordinate system  $(\mathbf{e}_1^{\dag}, \mathbf{e}_2^{\dag}, \mathbf{e}_3^{\dag}) $ positioned at the center of the Earth, with $\mathbf{e}_1^{\dag}$ pointing towards the North Pole, $\mathbf{e}_2^{\dag}$ pointing towards Null Island and $\mathbf{e}_3^{\dag}$ pointing to East. This ``new" Cartesian coordinate system can be thought of as a cyclical permutation of the classical Cartesian coordinate system at the Earth's center and on which classical spherical coordinates are based. See Appendix \ref{NS-Derivation} for a detailed description of both the classical and this rotated spherical coordinate system, and their relation.\\
Consider the set of the associated, right-handed, spherical coordinates $(\pd, \td, r')$, where  $\theta^{\dag}\in[-\frac{\pi}{2},\frac{\pi}{2}]$ and  $\varphi^{\dag}\in[0,2\pi)$ are the azimuthal and meridional angles, respectively, and $r'$ is the distance from Earth's center. The coordinates of a point $P$ on the Earth are, with respect to the coordinates $(\pd,\td,r')$,
\begin{equation}
r'\cos\td\cos\pd \ea^{\dag}+r'\cos\td\sin\pd \eb^{\dag}+r'\sin\td \ec^{\dag},
\end{equation}
and the North Pole has coordinates $\pd=\pi/2, \td=0, r'=R'$. Associated with the unit vectors $(\ep^{\dag}, \et^{\dag}, \er)$ in this $(\pd, \td, r')$-system are the velocity components $(u'_{\dag}, v'_{\dag}, w'_{\dag})$: $u'_{\dag}$ points from East to West, $v'_{\dag}$ from North to South and $w'_{\dag}$  points upward.\\
We use primes to represent dimensional (physical) variables. The prime will be removed once we introduce the appropriate set of non-dimensional variables.\\
In the new rotated spherical coordinates the momentum equations are
\begin{equation}\label{NS rotated1}
\begin{aligned}
& \rho' \frac{D} {D t'}\begin{pmatrix}
u'_{\dag} \\
v'_{\dag}\\
w'_{\dag}\\
\end{pmatrix}+\frac{\rho'}{r'}\begin{pmatrix}-u'_{\dag}v'_{\dag}\tan \theta^{\dag} + u'_{\dag}w'_{\dag}\\
u'^2_{\dag}\tan\theta^{\dag}+v'_{\dag}w'_{\dag}\\
-u'^2_{\dag}-v'^2_{\dag}
\end{pmatrix}+\\
&\quad +2\rho'\Omega'\begin{pmatrix}
    -v'_{\dag}\sin\varphi^{\dag}\cos\theta^{\dag}-w'_{\dag}\sin\varphi^{\dag}\sin\theta^{\dag}\\
    u'_{\dag}\sin\varphi^{\dag}\cos\theta^{\dag}-w'_{\dag}\cos\varphi^{\dag}\\
   u'_{\dag} \sin\varphi^{\dag}\sin\theta^{\dag} +v'_{\dag} \cos\varphi^{\dag}
\end{pmatrix}+\\
&\quad +\rho' r' \Omega'^2\begin{pmatrix}
    \sin\varphi^{\dag}\cos\varphi^{\dag}\cos\theta^{\dag}\\
    -\sin^2\varphi^{\dag} \sin\theta^{\dag}\cos\theta^{\dag}\\
   -\cos^2\varphi^{\dag}\cos^2\theta^{\dag} -\sin^2\theta^{\dag}
\end{pmatrix} = \\
& =-\nabla'p' +\rho\begin{pmatrix}
    0\\
    0\\
     -g'\frac{R'^2}{ r'^2}\\
\end{pmatrix}
+\Delta'_{\mu}
\begin{pmatrix}
    u'_{\dag}\\ v'_{\dag}\\ w'_{\dag}
\end{pmatrix}-\frac{1}{3}\begin{pmatrix}
    \frac{1}{r'\cos \theta^{\dag}}\frac{\partial}{\partial \varphi^{\dag}}\left(\frac{\mu'_H}{\rho'} \frac{D\rho'}{D t'}\right)\\
    \frac{1}{r'}\frac{\partial}{\partial \theta^{\dag}}\left(\frac{\mu'_H}{\rho'} \frac{D\rho'}{D t'}\right)\\
    \frac{\partial}{\partial r'}\left(\frac{\mu'_V}{\rho'} \frac{D\rho'}{D t'}\right)
\end{pmatrix}-\\
&\quad  -\frac{1}{r'^2\cos^2\theta^{\dag}}\begin{pmatrix}
    \mu'_H u'_{\dag}\\
    \mu'_H v'_{\dag}\\
    2\mu'_V(w'_{\dag}\cos^2\theta^{\dag}-v'_{\dag}\sin\theta^{\dag}\cos\theta^{\dag})
\end{pmatrix}+\frac{2\mu'_H}{r'^2}\frac{\partial}{\partial \theta^{\dag}}\begin{pmatrix}
0\\
 w'_{\dag}\\
   -v'_{\dag}
\end{pmatrix}+\\
&\quad  + \frac{2\mu'_H}{r'^2\cos\theta^{\dag}}\frac{\partial}{\partial \varphi^{\dag}}\begin{pmatrix}
w'_{\dag}-v'_{\dag}\tan\theta^{\dag}\\
 u'_{\dag}\tan\theta^{\dag}\\
   -u'_{\dag}
\end{pmatrix}+ \frac{d \mu'_V}{dr'}r' \begin{pmatrix}
    \frac{\partial}{\partial r'}\left(\frac{u'_{\dag}}{r'}\right)\\
       \frac{\partial}{\partial r'}\left(\frac{v'_{\dag}}{r'}\right)\\
        {0}
\end{pmatrix}+\\
&\quad +\frac{d \mu'_H}{d r'} \begin{pmatrix}
    \frac{1}{r'\cos \theta}\frac{\partial  {w'_{\dag}}}{\partial \varphi^{\dag}}\\
     \frac{1}{r'} \frac{\partial {w'_{\dag}}}{\partial \theta^{\dag}}\\
{0}
\end{pmatrix}+ \frac{d \mu'_V}{dr'} \begin{pmatrix}
   0\\
   0\\
   2 \frac{\partial w'_{\dag}}{\partial r'}  
\end{pmatrix}\\
&\quad  +\frac{d \mu'_V}{dr'}\begin{pmatrix}
   0\\
   0\\
   \frac{1}{r' \cos \theta^{\dag}} \frac{\partial u'_{\dag}}{\partial \varphi^{\dag}} + \frac{1}{r' \cos \theta^{\dag}} \frac{\partial}{\partial \theta^{\dag}} \left( v'_{\dag} \cos \theta^{\dag}  \right) + \frac{1}{r'^2} \frac{\partial}{\partial r'} \left( r'^2 w'_{\dag} \right) 
\end{pmatrix},
\end{aligned}
\end{equation}
where 
\begin{equation}
	\begin{aligned}
		\nabla'&= \left( \frac{1}{r'\cos\theta^{\dag}}\frac{\partial}{\partial\varphi^{\dag}},\  \frac{1}{r'}\frac{\partial}{\partial\theta^{\dag}} ,\   \frac{\partial}{\partial r'}
		\right),\\
		\frac{D}{Dt'}&=\frac{\partial} {\partial t'}+ \frac{u'_{\dag}}{r' \cos \theta^{\dag}} \frac{\partial }{\partial \varphi^{\dag}} + \frac{v'_{\dag}}{r'} \frac{\partial }{\partial \theta^{\dag}} + w'_{\dag}\frac{\partial }{\partial r'}, \\
		\Delta'_{\mu}&=\mu'_V \left(\frac{\partial^2}{\partial r'^2}+ \frac{2}{r'}\frac{\partial}{\partial r'}\right) + \frac{\mu'_H}{r'^2}\left(\frac{1}{ \cos^2 \theta^{\dag}} \frac{\partial^2 }{\partial \varphi^{\dag 2}}+\frac{\partial^2}{\partial \theta^{\dag 2}}   -  \tan\theta^{\dag}\frac{\partial}{\partial \theta^{\dag}}\right),
	\end{aligned}
\end{equation}
while the continuity equation is
\begin{equation}\label{mass rot1}
\frac{D'\rho}{Dt'}+\rho'\left[\frac{1}{r' \cos \theta^{\dag}} \frac{\partial u'_{\dag}}{\partial \varphi^{\dag}} + \frac{1}{r' \cos \theta^{\dag}} \frac{\partial}{\partial \theta^{\dag}} \left( v'_{\dag} \cos \theta^{\dag}  \right) + \frac{1}{r'^2} \frac{\partial}{\partial r'} \left( r'^2 w'_{\dag}\right)\right]=0;
\end{equation}
$R' \approx 6371\ km$ is the Earth's radius, $g' \approx 9.81 \ m\, s^{-2}$ is the average acceleration of gravity at the surface of the Earth, $\rho'$ is the fluid density,  $\mu'_V$ and $\mu'_H$ are the vertical and horizontal dynamic viscosities and  $\Omega' \approx 7.29\cdot10^{-5}\ s^{-1}$, is the angular velocity of Earth's rotation around the $\ea^{\dag}$-axis.\\
Moreover, we remark that even if the equation of state of seawater (derived from the first law of thermodynamics) is determined empirically in practice (see \cite{Roquet}), and  shows a complex relationship between density, temperature, and salinity, particularly in relatively shallow layers (see \cite{Talley}), under adiabatic conditions and assuming near-conservation of salinity, it simplifies to the equation of mass conservation (see \cite{CavalliniC}):
\begin{equation}\label{17}
	\frac{D\rho'} {D t'}=0.
\end{equation}
Although the thermohaline structure of the Arctic Ocean is intricate, the large excess of freshwater (resulting from direct precipitation and significant river runoff) forms a low-density surface layer that significantly reduces heat flux (see \cite{Meincke})\footnote{This runoff has been increasing as the hydrological cycle accelerates due to global warming \cite{Rawlins}. The increase in runoff is already measurable, with a reported $9.8\%$ rise over the 30-year period from 1977 to 2007 \cite{Overeem}.}. Since changes in salinity in the upper Arctic Ocean (up to about $5\%$) are generally seasonal (see \cite{Wadhams}), and thus negligible over the short time scale of days, we can adopt the mass conservation condition from equation \eqref{17}.\\
Moreover, assuming that the density of the fluid varies only with depth, namely $\rho'=\rho'(z)$, and that the dynamic viscosities are constant, namely $ \frac{	d \mu'_V}{dr}=\frac{	d \mu'_H}{dr}=0$,
we have that \eqref{NS rotated1} reduces to (up to dividing by $\rho'$):
\begin{equation}\label{govNS}
	\begin{aligned}
		&  \frac{D} {D t'}\begin{pmatrix}
			u'_{\dag} \\
			v'_{\dag}\\
			w'_{\dag}\\
		\end{pmatrix}+\frac{1}{r'}\begin{pmatrix}-u'_{\dag}v'_{\dag}\tan \theta^{\dag} + u'_{\dag}w'_{\dag}\\
			u'^2_{\dag}\tan\theta^{\dag}+v'_{\dag}w'_{\dag}\\
			-u'^2_{\dag}-v'^2_{\dag}
		\end{pmatrix}+\\
		&\quad+2\Omega'\begin{pmatrix}
			-v'_{\dag}\sin\varphi^{\dag}\cos\theta^{\dag}-w'_{\dag}\sin\varphi^{\dag}\sin\theta^{\dag}\\
			u'_{\dag}\sin\varphi^{\dag}\cos\theta^{\dag}-w'_{\dag}\cos\varphi^{\dag}\\
			u'_{\dag} \sin\varphi^{\dag}\sin\theta^{\dag} +v'_{\dag} \cos\varphi^{\dag}
		\end{pmatrix}+\\
        &\quad+r' \Omega'^2\begin{pmatrix}
			\sin\varphi^{\dag}\cos\varphi^{\dag}\cos\theta^{\dag}\\
			-\sin^2\varphi^{\dag} \sin\theta^{\dag}\cos\theta^{\dag}\\
			-\cos^2\varphi^{\dag}\cos^2\theta^{\dag} -\sin^2\theta^{\dag}
		\end{pmatrix} = 
		-\frac{1}{\rho'}\nabla'p' +\begin{pmatrix}
			0\\
			0\\
			-g'\frac{R'^2}{ r'^2}\\
		\end{pmatrix}+\\
		&\quad+\frac{1}{\rho'}\Delta'_{\mu}
		\begin{pmatrix}
			u'_{\dag}\\ v'_{\dag}\\ w'_{\dag}
		\end{pmatrix}
        -\frac{1}{r'^2\cos^2\theta^{\dag}}\begin{pmatrix}
			A'_H u'_{\dag}\\
			A'_H v'_{\dag}\\
			2A'_V(w'_{\dag}\cos^2\theta^{\dag}-v'_{\dag}\sin\theta^{\dag}\cos\theta^{\dag})
		\end{pmatrix}+\\
	&\quad +\frac{2A'_H}{r'^2}\frac{\partial}{\partial \theta^{\dag}}\begin{pmatrix}
			0\\
			w'_{\dag}\\
			-v'_{\dag}
		\end{pmatrix}+ \frac{2A'_H}{r'^2\cos\theta^{\dag}}\frac{\partial}{\partial \varphi^{\dag}}\begin{pmatrix}
			w'_{\dag}-v'_{\dag}\tan\theta^{\dag}\\
			u'_{\dag}\tan\theta^{\dag}\\
			-u'_{\dag}
		\end{pmatrix},
	\end{aligned}
\end{equation}
where
\begin{equation}
A'_H(z)=\frac{\mu'_H}{\rho'(z)}\qquad \text{and}\qquad A'_V(z)=\frac{\mu'_V}{\rho'(z)}
\end{equation}
are the horizontal and vertical kinematic eddy viscosities, while the continuity equation \eqref{mass rot1} reduces to
\begin{equation}\label{cont}
\frac{1}{r' \cos \theta^{\dag}} \frac{\partial u'_{\dag}}{\partial \varphi^{\dag}} + \frac{1}{r' \cos \theta^{\dag}} \frac{\partial}{\partial \theta^{\dag}} \left( v'_{\dag} \cos \theta^{\dag}  \right) + \frac{1}{r'^2} \frac{\partial}{\partial r} \left( r'^2 w ^{\dag}\right)=0.	
\end{equation}

\subsection{The governing equations at leading order}
Asymptotic expansions allow for the extraction of the main structure of the governing equations. To perform them, it is necessary to first non-dimensionalize the governing equations using physical scales that are representative of the specific phenomena, thereby clarifying the relative magnitudes of the terms by introducing suitable parameters that facilitate the development of asymptotic expansions. This approach makes it mathematically possible to obtain order-of-magnitude estimates for the relative importance of various factors. To apply this method to arctic ice-drift currents, we define the relevant physical scales (see \cite{Constantin2022} and the references therein):
\begin{equation}\label{scales}
	\begin{aligned}
		&R'\approx 6370\ km\qquad&\text{Earth radius,}\\
		&H'\approx 50\ m\qquad &\text{mean depth of the arctic sub-surface layer,}\\
		&L'\approx 10\ km&\text{horizontal length scale,}\\
		&\Bar{\rho'}\approx 1030\ kg\, m^{-3}&\text{average density of the arctic ocean,}\\
		&U' \approx 0.1\ m\, s^{-1}&\text{horizontal velocity scale.}
	\end{aligned}
\end{equation}
The non-dimensional variables $t, z, u, v, w, \rho, p, A_H, A_V$ are defined by

\begin{equation}\label{scaling pressure}
	\begin{aligned}
&t'=\frac{L'}{U'}t,\qquad	p'=p'_{atm}+\Bar{\rho'}U'^2 p+\Bar{\rho'}g'H'\int_z^0\rho(s)ds,\\ 
&r'=R'+H' z=R'(1+\epsilon z),\qquad (u'_{\dag}, v'_{\dag}, w'_{\dag})=U'(u,v,\kappa w),\\
&\rho'=\Bar{\rho'}\rho, \qquad A'_H=U'H' A_H, \qquad A'_V=\frac{U'H'^2}{L'}A_V,
\end{aligned}
\end{equation}
while the horizontal spatial variables $x, y$ will be defined later using a tangent plane approximation.
The scaling \eqref{scaling pressure} gives $t'\approx 1\, \text{day}$, $A'_H\approx 5\, m^2\, s^{-1}$ and $A'_V\approx 0.025\, m^2\, s^{-1}$.
Moreover, we introduce two adimensional parameters
\begin{equation} 
\begin{aligned}
    &\epsilon=\frac{H'}{R'}\approx10^{-5}\qquad\text{thin-shell parameter},\\
    &\delta=\frac{H'}{L'}\approx5\cdot10^{-3}\qquad\text{aspect-ratio parameter},
\end{aligned}
\end{equation}
and, as vertical velocities are of the order of approximately $10^{-6}\ m\, s^{-1}$, while for the horizontal velocity scale $U' \approx 0.1\ m\, s^{-1}$  (see \cite{Constantin2022}), we get that, for consistency, $\kappa=\mathcal{O}(\epsilon)$ (see \cite{CJ2023, Johnson2024}). We choose
\begin{equation}\kappa=\epsilon\end{equation}
and lastly we define
\begin{equation}
f= \frac{2\Omega' L'}{U'}\approx 14.5, 
\end{equation}
which is the inverse of the Rossby number $Ro=\nicefrac{U'}{2\Omega'L'}$. Using the previous scaling,  \eqref{govNS} became
\begin{equation}\label{primo scaling}
	\begin{aligned}
		&  \left[ \frac{1}{L'}\frac{\partial} {\partial t}+ \frac{u}{R'(1+\epsilon z) \cos \theta^{\dag}} \frac{\partial }{\partial \varphi^{\dag}} + \frac{v}{R'(1+\epsilon z) } \frac{\partial }{\partial \theta^{\dag}} + \frac{\epsilon w}{H'}\frac{\partial }{\partial z} \right]\begin{pmatrix}
			u \\
		v\\
		\epsilon w\\
		\end{pmatrix}U'^2+\\
        &\quad+\frac{U'^2}{R'(1+\epsilon z)}\begin{pmatrix}-uv\tan \theta^{\dag} + \epsilon uw\\
			u^2\tan\theta^{\dag}+\epsilon vw\\
			-u^2-v^2
		\end{pmatrix}+\\
		&\quad+\frac{2\Omega'}{U'}U'^2\begin{pmatrix}
			-v\sin\varphi^{\dag}\cos\theta^{\dag}-\epsilon w\sin\varphi^{\dag}\sin\theta^{\dag}\\
			u \sin\varphi^{\dag}\cos\theta^{\dag}-\epsilon w \cos\varphi^{\dag}\\
			u  \sin\varphi^{\dag}\sin\theta^{\dag} +v  \cos\varphi^{\dag}
		\end{pmatrix}+\\
        &\quad+R'(1+\epsilon z) \Omega'^2\begin{pmatrix}
			\sin\varphi^{\dag}\cos\varphi^{\dag}\cos\theta^{\dag}\\
			-\sin^2\varphi^{\dag} \sin\theta^{\dag}\cos\theta^{\dag}\\
			-\cos^2\varphi^{\dag}\cos^2\theta^{\dag} -\sin^2\theta^{\dag}
		\end{pmatrix}=\\
        & =-\frac{1}{\Bar{\rho'}\rho}\begin{pmatrix}
			\frac{1}{R'(1+\epsilon z)\cos \theta^{\dag}}\frac{\partial (\Bar{\rho'}U'^2p)}{\partial \varphi^{\dag}}\\
			\frac{1}{R'(1+\epsilon z)}\frac{\partial (\Bar{\rho'}U'^2p)}{\partial \theta^{\dag}}\\
		\frac{1}{H'}	\frac{\partial (\Bar{\rho'}g'H'\int_z^0\rho(s)ds+\Bar{\rho'}U'^2 p)}{\partial z}
		\end{pmatrix} -\\
	&\quad-\begin{pmatrix}
			0\\
			0\\
			g'\frac{1}{(1+\epsilon z)^2 }\\
		\end{pmatrix}	+\frac{U'^2}{L'}A_V \left(\frac{\partial^2}{\partial z^2}+ \frac{2\epsilon}{(1+\epsilon z)}\frac{\partial}{\partial z}\right)\begin{pmatrix}
		u \\
		v\\
		\epsilon w\\
	\end{pmatrix} +\\
    &\quad+\frac{U'^2 H'\, A_H}{(1+\epsilon z)^2}\left(\frac{1}{ R'^2\cos^2 \theta^{\dag}} \frac{\partial^2 }{\partial \varphi^{\dag 2}}+\frac{1}{R'^2}\frac{\partial^2}{\partial \theta^{\dag 2}}   -  \frac{\tan\theta^{\dag}}{R'^2}\frac{\partial}{\partial \theta^{\dag}}\right)\begin{pmatrix}
	u \\
	v\\
	\epsilon w\\
\end{pmatrix}-\\
	&\quad-\frac{U'^2 H'}{R'^2(1+\epsilon z)^2\cos^2\theta^{\dag}}\begin{pmatrix}
			A_H u\\
			A_H v\\
			2\delta A_V(\epsilon w\cos^2\theta^{\dag}-v\sin\theta^{\dag}\cos\theta^{\dag})
		\end{pmatrix}+\\
		& \quad+\frac{2U'^2 H'}{R'^2(1+\epsilon z)^2}A_H\frac{\partial}{\partial \theta^{\dag}}\begin{pmatrix}
			0\\
		\epsilon w\\
			-v
		\end{pmatrix}+ \\
        &\quad +\frac{2U'^2 H'}{R'^2(1+\epsilon z)\cos\theta^{\dag}}A_H\frac{\partial}{\partial \varphi^{\dag}}\begin{pmatrix}
			\epsilon w -v \tan\theta^{\dag}\\
			u \tan\theta^{\dag}\\
			-u 
		\end{pmatrix}.
	\end{aligned}
\end{equation}
In \eqref{primo scaling}, a horizontal scaling factor is missing. To address this, we introduce a tangent plane approximation given by:
\begin{equation}\label{tg}
	\left\{ 
	\begin{array}{ll}	
		x' = R' \cos\theta^{\dag}_0 (\varphi^{\dag} - \varphi^{\dag}_0), \\
		y' = R' (\theta^{\dag} - \theta^{\dag}_0),
	\end{array}
	\right.
\end{equation}
centered around the North Pole, where the coordinates are $\varphi_0^{\dag} = \pi/2$ and $\theta_0^{\dag} = 0$, and the scaling
\begin{equation}\label{scaling orizz}
	x' = L'x, \qquad y' = L'y.
\end{equation}
Using \eqref{tg} and \eqref{scaling orizz}, it follows that in the tangent plane approximation centered at the North Pole ($\varphi_0^{\dag} = \pi/2$, $\theta_0^{\dag} = 0$), the following transformations hold:
\begin{equation}\label{added tg}
	\frac{1}{R' \cos \theta^{\dagger}} \frac{\partial}{\partial \varphi^{\dagger}} \rightarrow \frac{1}{L'} \frac{\partial}{\partial x}, \qquad 
	\frac{1}{R'} \frac{\partial}{\partial \theta^{\dagger}} \rightarrow \frac{1}{L'} \frac{\partial}{\partial y}, \qquad 
	\frac{\tan \theta^{\dagger}}{R'} \frac{\partial}{\partial \theta^{\dagger}} \rightarrow 0.
\end{equation}
Using \eqref{tg}, \eqref{scaling orizz} and \eqref{added tg} in \eqref{primo scaling}, and dividing by $U'^2/L'$ both sides leads to
\begin{equation}\label{terzo scaling}
	\begin{aligned}
		&  \left[ \frac{\partial} {\partial t}+ \frac{u}{(1+\epsilon z) } \frac{\partial }{\partial x} + \frac{v}{(1+\epsilon z) } \frac{\partial }{\partial y} + \frac{\epsilon }{\delta}w\frac{\partial }{\partial z} \right]\begin{pmatrix}
			u \\
			v\\
			\epsilon w\\
		\end{pmatrix}+\\
    &\quad+   \frac{\epsilon}{\delta}\frac{1}{(1+\epsilon z)}\begin{pmatrix} \epsilon uw\\
			\epsilon vw\\
			-u^2-v^2
		\end{pmatrix}+f\begin{pmatrix}
			-v\\
			u \\
			0
		\end{pmatrix}=\\
        &=-\frac{1}{\rho} \begin{pmatrix}
			\frac{1}{(1+\epsilon z)}\frac{\partial p}{\partial x}\\
			\frac{1}{(1+\epsilon z)}\frac{\partial p}{\partial y}\\
			\frac{1}{\delta}\frac{\partial p}{\partial z}
		\end{pmatrix}+\begin{pmatrix}
			0\\
			0\\
		\frac{\epsilon}{\delta}\frac{(2z+\epsilon z^2)}{(1+\epsilon z)}\frac{1}{Fr^2}
		\end{pmatrix}+\\
		&	\quad +A_V \left(\frac{\partial^2}{\partial z^2}+ \frac{2\epsilon}{(1+\epsilon z)}\frac{\partial}{\partial z}\right)\begin{pmatrix}
			u \\
			v\\
			\epsilon w\\
		\end{pmatrix} + \frac{\delta}{(1+\epsilon z)^2}A_H\left( \frac{\partial^2 }{\partial x^2}+\frac{\partial^2}{\partial y^2} \right)\begin{pmatrix}
			u \\
			v\\
			\epsilon w\\
		\end{pmatrix}-\\
		&\quad-\frac{\epsilon^2}{\delta(1+\epsilon z)^2}\begin{pmatrix}
			A_H u\\
			A_H v\\
			2\epsilon\delta A_Vw
		\end{pmatrix}+2\frac{\epsilon}{(1+\epsilon z)^2}A_H\begin{pmatrix}
			\epsilon \frac{\partial w}{\partial x}\\
			\epsilon \frac{\partial w}{\partial y}\\
			- \frac{\partial u}{\partial x}- \frac{\partial v}{\partial y}
		\end{pmatrix},
	\end{aligned}
\end{equation} 
where $Fr=\frac{U'}{\sqrt{g'H'}}\approx 0.005$ is the Froude number, relating the inertia forces in a system to the effects of gravity.\\
Looking at the leading order terms in the three components of \eqref{terzo scaling} we get the following momentum equations:
\begin{equation}\label{governing enqs}
	\left\{ \begin{aligned}	
&	\frac{\partial u}{\partial t} + u \frac{\partial u}{\partial x} + v\frac{\partial u}{\partial y} - fv= -\frac{1}{\rho}\frac{\partial p}{\partial x} + A_V \frac{\partial^2 u}{\partial z^2},\\
	&\frac{\partial v}{\partial t} + u \frac{\partial v}{\partial x} + v\frac{\partial v}{\partial y} +fu= -\frac{1}{\rho}\frac{\partial p}{\partial y} + A_V \frac{\partial^2 v}{\partial z^2},\\
&\frac{\partial p}{\partial z}=0.
	\end{aligned}\right.
\end{equation}
In an analogous way, applying the scaling \eqref{scales} and \eqref{scaling pressure}, as well as the tangent plane approximation \eqref{added tg} to the continuity equation \eqref{cont} we get
\begin{equation}\label{scaled continuity 1}
	\frac{1}{1+\epsilon z}\left(\frac{\partial u}{\partial x}+\frac{\partial v}{\partial y}\right) + \frac{\epsilon}{\delta(1+\epsilon z)^2}\frac{\partial w}{\partial z}	=0,
\end{equation}
which, at leading order, reduces to
\begin{equation}\label{scaled cont}
	\frac{\partial u}{\partial x} + 	\frac{\partial v}{\partial y}=0.
\end{equation}
\subsection{Boundary conditions}
Let us now describe the boundary conditions associated to the ice-drift solutions to \eqref{governing enqs} and \eqref{scaled cont}.\\
Firstly, the velocity field $\bu'_{\dag}=(u'_{\dag}, v'_{\dag}, w'_{\dag})$ should posses a strong decay with depth, as ice-drift flows are negligible beneath the low-salinity surface layer (not in excess of $200\,m$). As the vertical velocity component is neglected at leading order in \eqref{governing enqs} and \eqref{scaled cont}, the boundary conditions capturing the decay with depth of wind drift flows is given by
\begin{equation}\label{21}
	\frac{	\partial (u'^2+v'^2)}{\partial z'}<0\quad\text{for}\quad z'<0.
\end{equation}\\
Additionally, we impose on the free surface the following condition on the pressure:
\begin{equation}\label{7}
p'=p'_{atm}\quad\text{on}\quad z'=0,
\end{equation}
where $p'_{atm}$ is the constant atmospheric pressure. The condition \eqref{7}, due to \eqref{scaling pressure} became
\begin{equation}\label{7new}
	p=0\quad\text{on}\quad z=0.
\end{equation}
The presence of the sea-ice cover, providing a resistive force against the motion forced by the wind and therefore damping surface waves, justifies the modeling assumption of a flat free surface $z'=0$, leading to the kinematic boundary condition
\begin{equation}\label{8}
	w'_{\dag}=0\quad\text{on}\quad z'=0,
\end{equation}
representing the absence of mass flux across the free boundary.\\
The continuity of the stress across the interface gives the surface boundary condition
\begin{equation}\label{3}
	(\tau^x,\tau^y)=\underbrace{\rho'(0)A'_V \left(\frac{\partial u'}{\partial z'},\frac{\partial v'}{\partial z'}\right)}_{\text{shear stress}}\quad\text{on}\quad z'=0,
	\end{equation}
relating the wind-ice-stress vector 	$(\tau^x,\tau^y)$ with the shear stress. \\
The wind-ice-water stress can be partitioned, in general, as (see \cite{Constantin2022})
\begin{equation}\label{4}
	(\tau^x,\tau^y)=\alpha \boldsymbol{\tau}_{\mathrm{\mathrm{ice}}}+(1-\alpha) \boldsymbol{\tau}_{\mathrm{air}},
\end{equation}
where $\alpha$ is the ice-covered surface area fraction, $\boldsymbol{\tau}_{\mathrm{ice}}$ is the ice-water stress and $\boldsymbol{\tau}_{\mathrm{air}}$ is the air-water stress. As we are considering flows in a region around the North Pole, the surface is always covered with ice ($\alpha=1$), hence equation \eqref{4} reduces to 
\begin{equation}\label{4 bis}
	(\tau^x,\tau^y)=\boldsymbol{\tau}_{\mathrm{ice}}.
\end{equation}
The ice-water stress $\boldsymbol{\tau}_{\mathrm{ice}}$ is given by the bulk formula 
(see \cite{Constantin2022} or \cite{Yang})
\begin{equation}\label{6}
\boldsymbol{\tau}_{\mathrm{ice}}=\rho'(0) C_{\mathrm{ice}}|\mathbf{U}'_{\mathrm{ice}}-\mathbf{U}'|(\mathbf{U}'_{\mathrm{ice}}-\mathbf{U}'),
\end{equation}
where $\rho'(0)\approx 1026\ kg\, m^{-3}$ is the water density at the surface, $C_{\mathrm{ice}}\approx 5.5\cdot10^{-3}$ is the ice-water drag coefficient, $\mathbf{U}'_{\mathrm{ice}}$ is the ice velocity and $\mathbf{U}'$ is the surface current velocity.\\
Combining \eqref{3} and \eqref{6} gives the following boundary condition describing the horizontal momentum exchange at the surface:
\begin{equation}\label{22}
	\rho'(0)A_V'\left(\frac{\partial u'}{\partial z'},\frac{\partial v'}{\partial z'}\right)=\rho'(0) C_{\mathrm{ice}} |\mathbf{U}'_{\mathrm{ice}}-\mathbf{U}'|(\mathbf{U}'_{\mathrm{ice}}-\mathbf{U}')
	\quad\text{on}\quad z'=0,
\end{equation}
which becomes
\begin{equation}\label{stress adim}
\left(\frac{\partial u}{\partial z},\frac{\partial v}{\partial z}\right)=\frac{C_i}{A_V} |(u_i-u,v_i-v)|(u_i-u,v_i-v) \quad\text{on}\quad z=0,
\end{equation} after non-dimensionalizing with 
\begin{equation}
 \mathbf{U}'_{\mathrm{ice}}=U'(u_i,v_i)\quad\text{and}\quad
C_{\mathrm{ice}}=\frac{H'}{L'}C_i.
\end{equation}
Summing up, the boundary conditions in non-dimensional form relevant to the nonlinear drift flow are
\begin{equation}\label{bd adim}
	\begin{cases}
p=0&\text{on}\   z=0,\\
\frac{	\partial (u^2+v^2)}{\partial z}<0&\text{for}\  z<0,\\
\left(\frac{\partial u}{\partial z},\frac{\partial v}{\partial z}\right)=\frac{C_i}{A_V} |(u_i-u,v_i-v)|(u_i-u,v_i-v) &\text{on}\ z=0.
\end{cases}
\end{equation}
The system \eqref{governing enqs} and \eqref{scaled cont}, with its simple $z$-dependence, allows for solutions that involve layered, depth-dependent $2D$ incompressible horizontal flows. This will be examined in more detail in the following section, adopting a Lagrangian formulation.
\section{The nonlinear ice-drift solution in Lagrangian framework}\label{sec-Lag}
\begin{figure}\label{TIKZ1}
\centering
	\begin{tikzpicture}
	
	\draw[line width=.3mm] (0,0) circle (3);
	
	\node at (0, 3.5) {180};
	\node at (0, -3.5) {0};
	\node at (3.5, 0) {90 E};
	\node at (-3.5, 0) {90 W};
	
\draw[->, blue, line width=1.5mm] (80:1.7) -- (260:1.7);
		\node at (75:2.1) {\textcolor{blue}{TDC}};
	\draw[->, ultra thick] (0, 0) -- (-1, 0);
	\draw[->, ultra thick] (0, 0) -- (0, -1);
	
	\node at (-1, 0.3) {$\ep^{\dag}$};
	\node at (0.3, -1) {$\et^{\dag}$};
	
	\foreach \i in {0,30,...,330} {
		\draw[-, ultra thin, dashed] (0, 0) -- (\i:3);
	}
\end{tikzpicture}
\caption{The base vector $\ep^{\dag}, \et^{\dag}$ of the rotated spherical coordinate system at the North Pole, and a schematic depiction of the geostrophic component of the Transpolar Drift Current (TDC)}
\end{figure}
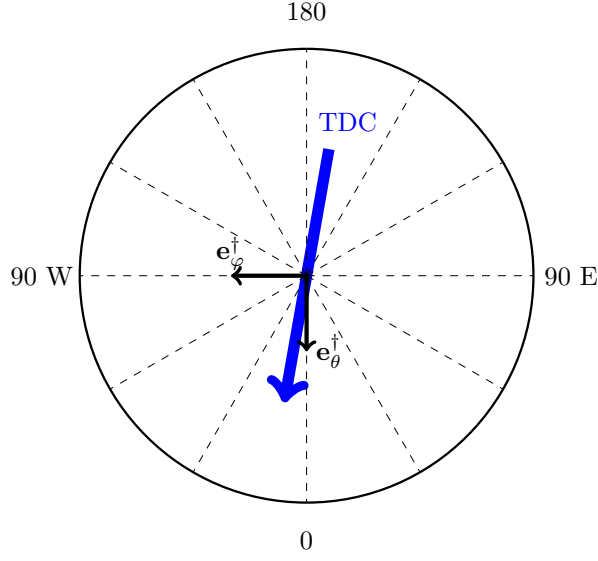

Typically, the motion of a fluid is studied by adopting the so-called Eulerian approach, in which the motion of a fluid is described by observing it from a fixed reference frame in space. In this approach, the focus is on analyzing how the properties of the fluid (such as velocity, pressure, density, temperature, etc.) change over time at specific points in space, rather than tracking the motion of individual fluid particles \cite{Kundu}.\\
There is another approach to describe the motion of a fluid, named Lagrangian, in which one follows the motion of fixed fluid particles, whereby the coordinates of such an individual particle, $x,y,z$ are considered to depend on three spatial coordinates, $\mathsf{a}, \mathsf{b}, \mathsf{c}$, and time $t$, called Lagrangian coordinates or material variables or labeling variables (see \cite{AbrashkinPelinovsky}).\\
The central part of the Arctic Ocean features the presence of the Transpolar Drift Current (TDC), which at the North Pole is steady and will be modeled as the sum of an Ekman-type spiral plus a geostrophic flow (see \cite{CJ2024}). In our model we assume the amplitude of the Ekman-type component of the TDC to be independent of $x$ and $y$ and $\mathfrak{u}_g$ and $\mathfrak{v}_g$ representing the geostrophic flow (which is assumed to be constant for simplicity). Therefore, the Transpolar Drift current can be modeled as the velocity vector
\begin{equation}\label{TDC}
    \text{TDC}=\left(m e^{sz}\cos(sz+\alpha)+\mathfrak{u}_g,\  m e^{sz}\sin(sz+\alpha)+\mathfrak{v}_g\right),
\end{equation}
where $m$ is the amplitude of the Ekman-type spiral\footnote{We call ``Ekman-type" spiral (or flow or current) a velocity vector of the form $e^{(1+i)sz}$ where $s\in\mathbb{R}^+\setminus\{\lambda\}=\{\tilde{s}\in \mathbb{R}\,:\, \tilde{s}>0\}\setminus\{\lambda\}$, while we call Ekman spiral (or flow or current) velocity vector of the form $e^{(1+i)\lambda z}$, where $\lambda=\sqrt{\frac{f}{2 A_V}}$.} of the TDC at the surface, $s$ is a multiplication factor representing ``how fast" the Ekman-type flow should decay to zero, $\alpha$ is the angle that represents the direction of the TDC at the surface and finally $\mathfrak{u}_g$ and $\mathfrak{v}_g$ are the components of the constant geostrophic flow. See Figure \ref{TIKZ1} for an approximate depiction of the geostrophic component of the Transpolar Drift Current. As the geostrophic component of the Transpolar Drift Current should be almost parallel to the Greenwich Meridian (the meridian of $0^{\circ}$ latitude), we must require $\mathfrak{v}_g>\mathfrak{u}_g$. \\
Adopting the Lagrangian approach, and specifying at  every time $t$ the positions
\begin{equation}\label{Lag}
	\left\{ \begin{aligned}	
		x(t; a,b,z)=a + d_1(z)t -\frac{1}{k} e ^{k(b+z)}\sin(k(a-z-ct)) + \\
        +{\left(m e^{sz}\cos(sz+\alpha)+\mathfrak{u}_g\right)t}\\
			y(t; a,b,z)=b + d_2(z)t +\frac{1}{k} e ^{k(b+z)}\cos(k(a-z-ct)) +\\
           +\left(  m e^{sz}\sin(sz+\alpha)+\mathfrak{v}_g\right)t
	\end{aligned}\right.
\end{equation}
of the horizontally moving fluid particles in terms of the depth $z$, the material variables $a,b$ and the parameters $k>0$ and $c>0$, for suitably chosen functions $d_1(z)$ and $d_2(z)$, we will show that a solution to the nonlinear ice-drift flow at the North Pole \eqref{governing enqs}, \eqref{scaled cont} satisfying the boundary conditions \eqref{bd adim} is given by \eqref{Lag}. The time-periodic functions in \eqref{Lag} have principal time-period $T=\frac{2\pi}{ck}$.\\
The material variable \(a\) is assumed to be a real number, while \(b\) must satisfy \(b_1 < b < b_2 < 0\), ensuring that for \(z \leq 0\), the horizontal oscillations in \eqref{Lag} of a particle decay exponentially with increasing depth. \\
We anticipate that, due to the structure of $d(z)=(d_1(z),\, d_2(z))$, (cf. \eqref{38} and \eqref{44}), the position at every time $t$ will be
\begin{equation}\label{postion}
    \left\{
    \begin{aligned}	
        x(t; a,b,z) = a + \left[\left(d_1(0)+ m \cos(\alpha)\right)e^{(1+i)\lambda z}+\mathfrak{u}_g\right]t -\\
  -\frac{1}{k} e ^{k(b+z)}\sin(k(a-z-ct)), \\
        y(t; a,b,z) = b + \left[\left(d_2(0)+  m \sin(\alpha)\right)e^{(1+i)\lambda z}+\mathfrak{v}_g\right]t  +\\
            +\frac{1}{k} e ^{k(b+z)}\cos(k(a-z-ct)) .
    \end{aligned}
    \right.
\end{equation}
where $\lambda=\sqrt{\frac{f}{2 A_V}}$ is the inverse of the Ekman depth (see \cite{Kundu}). Namely, at each vertical level \(z\) beneath the ocean surface (\(z = 0\)), \eqref{postion} represents a depth-dependent oscillatory motion, characterized by a varying amplitude and phase shift, superimposed on a Ekman current propagating in the direction of the horizontal vector \(\left([d_1(0)+ m\cos(\alpha)]e^{(1+i)\lambda z},\ [d_2(0)+ m \sin(\alpha)]e^{(1+i)\lambda z} \right)\) and on a constant geostrophic flow $(\mathfrak{u}_g,\mathfrak{v}_g)$.\\
For simplicity we write
\begin{equation}\xi=k(a-z-ct),\end{equation}
and it follows that, at every fixed time $t$, 
\begin{equation}\label{jac1}
\begin{pmatrix}
	\frac{	\partial x}{\partial a} & \frac{	\partial x}{\partial b}\\
	\frac{	\partial y}{\partial a} & \frac{	\partial y}{\partial b}
\end{pmatrix}=\begin{pmatrix}
1-e^{k(b+z)}\cos(\xi) & -e^{k(b+z)}\sin(\xi)\\
-e^{k(b+z)}\sin(\xi) &1+e^{k(b+z)}\cos(\xi)
\end{pmatrix}.
\end{equation}
The determinant of the matrix in \eqref{jac1} is equal to $1-e^{2k(b+z)}$, and its inverse is
\begin{equation}
	\begin{pmatrix}
		\frac{	\partial a}{\partial x} & \frac{	\partial a}{\partial y}\\
		\frac{	\partial b}{\partial x} & \frac{	\partial b }{\partial y}
	\end{pmatrix}=\frac{1}{1-e^{2k(b+z)}}\begin{pmatrix}
		1+e^{k(b+z)}\cos(\xi) & e^{k(b+z)}\sin(\xi)\\
		e^{k(b+z)}\sin(\xi) &1-e^{k(b+z)}\cos(\xi)
	\end{pmatrix}.
\end{equation}
From \eqref{Lag} we have that
\begin{equation}\label{Lag u v}
	\left\{ \begin{aligned}	
	&u=\frac{\partial x}{\partial t}=\overbrace{d_1(z) +c e ^{k(b+z)}\cos(\xi)+m e^{sz}\cos(sz+\alpha)}^{:=\hat{u}}+\mathfrak{u}_g,\\
	&v=\frac{\partial y}{\partial t}=\underbrace{d_2(z) +ce ^{k(b+z)}\sin(\xi)+m e^{sz}\sin(sz+\alpha)}_{:=\hat{v}}+\mathfrak{v}_g.
	\end{aligned}\right.
\end{equation}
where with $\hat{u}, \hat{v}$ we indicate the non-geostrophic components of the horizontal velocities $(u,v)$, consisting of the Ekman flow plus oscillations. Moreover, since
\begin{equation}\label{Jac2}
\begin{aligned}
&\begin{pmatrix}
	\frac{	\partial u}{\partial x} & \frac{	\partial v}{\partial x}\\
	\frac{	\partial u}{\partial y} & \frac{	\partial v}{\partial y}
\end{pmatrix} =	\begin{pmatrix}
		\frac{	\partial a}{\partial x} & \frac{	\partial b}{\partial x}\\
		\frac{	\partial a}{\partial y} & \frac{	\partial b}{\partial y}
	\end{pmatrix} 
\begin{pmatrix}
		\frac{	\partial u}{\partial a} & \frac{	\partial v}{\partial a}\\
		\frac{	\partial u}{\partial b} & \frac{	\partial v}{\partial b}
	\end{pmatrix}= \\
&\ =\frac{kc}{1-e^{2k(b+z)}}\begin{pmatrix}
 -e^{k(b+z)}\sin(\xi)&  e^{k(b+z)}\cos(\xi) +  e^{2k(b+z)}\\
e^{k(b+z)}\cos(\xi)- e^{2k(b+z)} & e^{k(b+z)}\sin(\xi)
\end{pmatrix},    
\end{aligned}
\end{equation}
at each horizontal level the two-dimensional horizontal flow \eqref{Lag} has constant vorticity
\begin{equation}\label{vorticity}
	\frac{\partial u}{\partial y}- \frac{\partial v}{\partial x}=\frac{2kc }{1-e^{-2k(b+z)}}.
	\end{equation}
    Let us decompose the pressure as
    \begin{equation}
        p=P_G+\hat{p}
    \end{equation}
   where
    \begin{equation}\label{geostrophic}
    \left\{\begin{aligned}
\frac{\partial P_G}{\partial x}&=  \rho f \mathfrak{v}_g,\\
   \frac{\partial P_G}{\partial y}&=    -\rho f \mathfrak{u}_g.
     \end{aligned}\right.
       \end{equation}
   In general, pressure gradients are balanced by the geostrophic current (see \cite{Kundu}), that is to say that $\frac{\partial p}{\partial x}\approx \rho f \mathfrak{v}_g= \frac{\partial P_G}{\partial x} $ and $\frac{\partial p}{\partial y}\approx -\rho f \mathfrak{u}_g=\frac{\partial P_G}{\partial y} $ where the last equalities are due to \eqref{geostrophic}. For this reason we assume 
   \begin{equation}\label{phat}
   	\frac{\partial \hat{p}}{\partial x}=\frac{\partial \hat{p}}{\partial y}=0
   \end{equation}
As the first two equations in \eqref{governing enqs} are
\begin{equation}
	\left\{ \begin{aligned}	
		\frac{D u}{D t}  - fv= -\frac{1}{\rho}\frac{\partial p}{\partial x} + A_V \frac{\partial^2 u}{\partial z^2},\\
		\frac{D v}{D t}+fu= -\frac{1}{\rho}\frac{\partial p}{\partial y} + A_V \frac{\partial^2 v}{\partial z^2},
	\end{aligned}\right.
\end{equation}
using \eqref{geostrophic} we can write 
\begin{equation}
	\left\{ \begin{aligned}	
		\frac{D u}{D t}  - f\hat{v}= -\frac{1}{\rho}\frac{\partial \hat{p}}{\partial x} +
        A_V \frac{\partial^2 u}{\partial z^2},\\
		\frac{D v}{D t}+f\hat{u}= -\frac{1}{\rho}\frac{\partial \hat{p}}{\partial y} +
        A_V \frac{\partial^2 v}{\partial z^2}.
	\end{aligned}\right.
\end{equation}
which reduces to 
\begin{equation}\label{governing enqs 2}
	\left\{ \begin{aligned}	
		\frac{D u}{D t}  - f\hat{v}=		A_V \frac{\partial^2 u}{\partial z^2},\\
		\frac{D v}{D t}+f\hat{u}=		A_V \frac{\partial^2 v}{\partial z^2}.
	\end{aligned}\right.
\end{equation}
under the approximation \eqref{phat}.\\
Moreover, as in Lagrangian framework the material derivative corresponds to the time derivative, we have that
\begin{equation}\label{Lag du dv}
	\left\{ \begin{aligned}	
&\frac{D u}{D t}=kc^2 e ^{k(b+z)}\sin(\xi),\\
		&\frac{D v}{D t}=-kc^2e ^{k(b+z)}\cos(\xi),
	\end{aligned}\right.
\end{equation}
therefore inserting \eqref{Lag u v} and \eqref{Lag du dv} into \eqref{governing enqs 2} we get
\begin{equation}\label{32}
	\left\{ \begin{aligned}	
		0=\left[f d_2(z)+A_V d_1''(z)+(f-2A_Vs^2)me^{sz}\sin(sz+\alpha)\right]+\\
    +\left(fc + 2 A_V k^2 c - k c^2\right)e^{k(b+z)}\sin(\xi),\\
		0=-\left[f d_1(z)-A_V d_2''(z)+ (f-2A_Vs^2)me^{sz}\cos(sz+\alpha)\right]-\\
       -\left(fc + 2 A_V k^2 c - k c^2\right)e^{k(b+z)}\cos(\xi).
	\end{aligned}\right.
\end{equation}
The system \eqref{32} is satisfied if
\begin{equation}
	\left\{ \begin{aligned}	
		f d_2(z)+A_V d_1''(z)+(f-2A_Vs^2)me^{sz}\sin(sz+\alpha)=0,\\		
	-f d_1(z)+A_V d_2''(z)- (f-2A_Vs^2)me^{sz}\cos(sz+\alpha)=0,\\
		\left(fc + 2 A_V k^2 c - k c^2\right)e^{k(b+z)}=0.\\
	\end{aligned}\right.
\end{equation}
Therefore, given the dispersion relation 
\begin{equation}\label{34}
c=\frac{f}{k}+2A_V k,
\end{equation} 
the system \eqref{32} reduces to
\begin{equation}\label{32bis}
	\left\{ \begin{aligned}	
		&0= f d_2(z)+A_V d_1''(z)+(f-2A_Vs^2)me^{sz}\sin(sz+\alpha),\\
		&0 =-f d_1(z)+A_V d_2''(z)- (f-2A_Vs^2)me^{sz}\cos(sz+\alpha),
	\end{aligned}\right.
\end{equation}
and, writing in complex-variable notation,
\begin{equation}
d(z)=d_1(z) + i d_2(z)\qquad \text{and}\qquad \mathcal{K}=m\frac{f-2A_Vs^2}{A_V}=2m(\lambda^2-s^2),
\end{equation}the system \eqref{32bis} becomes
\begin{equation}
d''(z)-i\frac{f}{A_V} d(z)=i\mathcal{K}e^{i\alpha}e^{(1+i)sz}.
\end{equation}
If $s=\lambda$, $\mathcal{K}=0$,  the solution is given by
\begin{equation}
	d(z)=A_1 e^{(1+i)\lambda z}+ B_1e^{-(1+i)\lambda z}=A_1 e^{(1+i)\lambda z}+ B_1e^{-(1+i)\lambda z},
\end{equation}
where $\lambda=\sqrt{\frac{f}{2A_V}}$ and $A_1, B_1 \in \mathbb{C}$ are complex constants.  On the other hand, if  $s\not=\lambda$, given two complex constants $A_2, B_2 \in \mathbb{C}$, the solution is 
\begin{equation}
\begin{aligned}
    d(z)&=A_2 e^{(1+i)\lambda z}+ B_2e^{-(1+i)\lambda z}+ \frac{\mathcal{K}}{2(s^2-\lambda^2)}e^{i\alpha}e^{(1+i)sz}=\\
    &=A_2 e^{(1+i)\lambda z}+ B_2e^{-(1+i)\lambda z}-me^{i\alpha}e^{(1+i)sz}.
\end{aligned}	
\end{equation}
As wind drift currents are insignificant at great depth, we need to set $B_1=B_2=0$, and equaling the previous two equations for $d(z)$ at $z=0$ gives $A_1=A_2-me^{i\alpha}$, finally leading to \begin{equation}\label{38}
d(z)=(d(0)+me^{i\alpha}) e^{(1+i)\lambda z}-me^{i\alpha}e^{(1+i)sz},\qquad\forall\,s\in\mathbb{R}^+ ,
\end{equation}
with $d(0)=d_1(0)+id_2(0)\in \mathbb{C}$ being the the mean ice-drift current at $z=0$, and $(m\cos(\alpha)+\mathfrak{u}_g, m\sin(\alpha)+\mathfrak{v}_g)=:(\mathfrak{u}_{TDC}, \mathfrak{v}_{TDC})=:\mathfrak{U}_{TDC}$ representing the Transpolar Drift Current at the surface.\\
The time-average of the velocity \eqref{Lag u v}, which in view of \eqref{38} reads as
\begin{equation}  
\begin{aligned}
u+iv&=\overbrace{(d_1(0)+m\cos(\alpha))e^{(1+i)\lambda z}+i(d_2(0)+m\sin(\alpha))e^{(1+i)\lambda z}}^{d(z)=d_1(z)+id_2(z)}+\\
&\ +c e^{k(b+z)}e^{ik(a-z-ct)}+(\mathfrak{u}_g+i\mathfrak{v}_g),
\end{aligned}
\end{equation}
over a period $T=\frac{2\pi}{ck}=\frac{2\pi}{f+2A_Vk^2}$ yields the mean-drift current, which is, due to \eqref{38}, an Ekman spiral $d(z)$ plus the geostrophic component $\mathfrak{U}_{g}=\mathfrak{u}_g+i\mathfrak{v}_g$ of the TDC 
\begin{equation} \label{meancurrent}
\langle u+iv\rangle_T=(d_1(0)+id_2(0) +m\cos(\alpha)+i\,m\sin(\alpha))e^{(1+i)\lambda z}+\mathfrak{u}_g+i\mathfrak{v}_g=d(z)+\mathfrak{U}_{g},\end{equation} 
where $\langle\cdot\rangle_T$ represents the time-average.\\
Due to the structure of the mean current  \eqref{meancurrent}, the boundary condition \eqref{stress adim} reduces to 
\begin{equation}\label{40}
\frac{\lambda A_V}{C_i} \left(\D_1-\D_2, \D_1+\D_2\right)= |(u_i-\mathfrak{u}_g-\D_1,v_i-\mathfrak{v}_g-\D_2)|(u_i-\mathfrak{u}_g-\D_1 ,v_i-\mathfrak{v}_g-\D_2),
\end{equation}
in which the unknown vector components $\D_1=d_1(0)+m\cos(\alpha)$ and $\D_2=d_2(0)+m\sin(\alpha)$ represent the non-dimensional ice-drift non-geostrophic surface current. Writing 
\begin{equation}
R e^{i \theta}=(u_i-\mathfrak{u}_g-\D_1)+i(v_i-\mathfrak{v}_g-\D_2), 
\end{equation}
\eqref{40} can be rewritten as
\begin{equation}\label{41 previous}
	\beta\left[(\D_1-\D_2)+i( \D_1+\D_2)\right]=R^2 e^{i \theta},
\end{equation}
or equivalently, as
\begin{equation}\label{41}
	R^2 e^{i \theta}+\beta(1+i)	R e^{i \theta}=\zeta,
\end{equation}
with $\beta=\frac{\lambda A_V}{C_i}>0$, for the unknowns $R\geq 0$ and $\theta \in [0,2\pi)$, where
\begin{equation}
\zeta=\beta[(u_i-\mathfrak{u}_g-v_i+\mathfrak{v}_g)+i(u_i-\mathfrak{u}_g+v_i-\mathfrak{u}_g)] \in \mathbb{C}.
\end{equation}
As the ice moves faster than the Trasnpolar Drift Current (see \cite{CJ2024}), $\zeta\not=0$. We therefore infer from \eqref{41} that $R\not=0$ and 
\begin{equation}\label{42}
e^{i \theta}=\frac{	\zeta}{R[R+\beta(1+i)]}.
\end{equation}
Multiplying \eqref{42} by its complex conjugate gives the quartic equation
\begin{equation}\label{43}
\mathcal{P}(R)=R^4+2\beta R^3+2\beta^2R^2-|\zeta|^2=0,
\end{equation}
and, since $\mathcal{P}(0)<0$, $\lim_{R\rightarrow\pm \infty} \mathcal{P}(R)=+\infty$ and $\mathcal{P}''(R)>0$, $\mathcal{P}$ is strictly convex, ensuring that there is one solution $R>0$ of \eqref{43}.\\
We remark that our analysis shows that the boundary condition \eqref{stress adim} uniquely determines the surface current $d(0)$ knowing the sea-ice velocity and the TDC velocity $\mathfrak{U}_{TDC}$ at the surface.\\
Writing $(u_i-\mathfrak{u}_g)+i (v_i-\mathfrak{v}_g)=\mathcal{U}e^{i\phi}$ with $\phi\in[0,2\pi)$ and $\mathcal{U}>0$, and $\D_1+i\D_2 =\D e^{i\psi}$ with $\psi\in[0,2\pi)$ and $\D>0$, $R=|\mathcal{U}e^{i\phi}-\D e^{i\psi}|>0$ and \eqref{40} can be rewritten as
\begin{equation}\label{deflection}
\mathcal{U}e^{i(\phi-\psi)}=\D+\beta\frac{(1+i)\D}{|\mathcal{U}e^{i\phi}-\D e^{i\psi}|}.
\end{equation}
As the right-hand side of \eqref{deflection} is a complex number with real part bigger than the (positive) imaginary part, it follows that $0<\phi-\psi<\nicefrac{\pi}{4}$.\\
Writing the ice velocity  $\mathbf{U}_{\mathrm{ice}}=u_i+i\,v_i$ as
\begin{equation}
    \mathbf{U}_{\mathrm{ice}}=\mathfrak{U}_{TDC}+\mathcal{V}_{\mathrm{rel}}=\mathfrak{U}_{g}+(m\cos\alpha+i m\sin\alpha)+\mathcal{V}_{\mathrm{rel}},
\end{equation}
where $\mathcal{V}_{\mathrm{rel}}$ is the vector of the ice velocity relative to the TDC, we have that
\begin{equation}
  \mathcal{U}e^{i\phi}=  \mathbf{U}_{\mathrm{ice}}-\mathfrak{U}_{g}=(m\cos\alpha+i m\sin\alpha)+\mathcal{V}_{\mathrm{rel}}.
\end{equation}
As we have defined  $\mathcal{D}e^{i\psi}$ as
\begin{equation}
    \mathcal{D}e^{i\psi}= \left(d_1(0)+id_2(0)\right)+\left(m\cos(\alpha)+im\sin(\alpha)\right),
\end{equation}
the fact that $0<\phi-\psi<\nicefrac{\pi}{4}$ shows that the surface current vector $\left(d_1(0)+id_2(0)\right)$ is directed to the right of the ice moving relatively to the TDC, forming an angle of less than $45^{\circ}$ with its path.\\
Finally, observe that, for any given $k>0$, the horizontal particle path given by \eqref{Lag} coupled with \eqref{38} is
\begin{equation}\label{44}
\begin{aligned}
    x(t;a,b,z)+iy(t;a,b,z)=(a+ib)+\bigg[(d(0)+me^{i\alpha})e^{(1+i)\lambda z}+\\
    +\mathfrak{U}_{g}\bigg]t+ \frac{1}{k}e^{k(b+z)}e^{i\left(\frac{\pi}{2}+k(a-z)-(f+2A_Vk^2)t\right)},
\end{aligned}
\end{equation}
representing trochoids, and is a solution of \eqref{governing enqs} and \eqref{scaled cont}.\\
The mean-drift current, over a period $T=\frac{2\pi}{ck}=\frac{2\pi}{f+2A_Vk^2}$,
\begin{equation}\label{45}
	\begin{aligned}
\langle u+iv\rangle_T&=\frac{1}{T}\int_0^T\left( u(t;a,b,z)+iv(t;a,b,z)\right) dt=\\
&=(d(0)+me^{i\alpha})e^{(1+i)\lambda z}+\mathfrak{U}_{g}
	\end{aligned}
\end{equation}
is the superposition of a classical Ekman spiral and a geostrophic current, while the oscillatory perturbation of the current 
\begin{equation}
\frac{1}{k}e^{k(b+z)}e^{i\left(\frac{\pi}{2}+k(a-z)-(f+2A_Vk^2)t\right)}
\end{equation}
is essentially inertial, as  $A_V=\mathcal{O}(1)$ and $k<<1$ gives
\begin{equation}
f+2A_Vk^2\approx f
\end{equation}
for the frequency. Averaging over the depth the Ekman component of \eqref{45} gives the depth-averaged mean-drift Ekman  of the solution \eqref{44}, also known as (time-averaged) Ekman trasport,
\begin{equation}\label{48}
\mathcal{I}_{Ek}:=\int_{-\infty}^0 (d(0)+me^{i\alpha})e^{(1+i)\lambda z}\, dz=\frac{(d(0)+me^{i\alpha})}{\lambda \sqrt{2}}e^{-i\frac{\pi}{4}},
\end{equation}
which is at $45^{\circ}$ to the right of the Ekman surface current $d(0)+me^{i\alpha}$. Given that $d(0)+me^{i\alpha}$ is directed to the right of $\mathbf{U}_{\mathrm{ice}}-\mathfrak{U}_g$ (due to \eqref{deflection}), as a consequence we have that $\mathcal{I}_{Ek}$ is directed to the right of  $\mathbf{U}_{\mathrm{ice}}-\mathfrak{U}_g$ with an angle between $45^{\circ}$ and $90^{\circ}$.\\
Observe that, for every fixed value of $z$, the particle path \eqref{44} represents a trochoid, the only known solution for the $2D$ incompressible Euler equations with free boundary discovered by Gerstner (see \cite{Milne} and \cite{AbrashkinPelinovsky}). Therefore, the solution \eqref{44} of the governing equation for arctic ice-drift flow represents a superposition of an Ekman spiral, a Gerstner-type solution and a geostrophic current.
\section{Discussion}\label{Sec-Discussion}
Our analysis provided an explicit solution to the nonlinear governing equations describing the leading-order dynamics of ice-drift flows in a region centered around the North Pole. Such governing equations have been obtained from the Navier-Stokes and continuity equations adopting a thin-shell approximation coupled with a tangent plane approximation. The solution, in the Lagrangian formalism, highlights the superposition of near-inertial oscillations on a mean flow, the latter being the sum of an Ekman spiral (partially driven by the Transpolar Drift Current's Ekman-type spiral and partially driven by ice stress) and of the geostrophic component of the TDC. This solution extends the one found in \cite{Constantin2022} and \cite{Constantin2022NOTE} for regions outside the Amundsen Basin, where the Transpolar Drift Current is not present:
    \begin{equation}  \label{eqC2022}
\tilde{u}+i\tilde{v}=(\Tilde{d}_1(0)+i\tilde{d}_2(0))e^{(1+i)\lambda z}+c e^{k(b+z)}e^{ik(a-z-ct)}.
\end{equation}
In \eqref{eqC2022}, the entirety of the ``forcing" velocity (namely the velocity of ice or that of the wind, or a combination of them, as regions partially covered by ice where considered in \cite{Constantin2022} and \cite{Constantin2022NOTE}) induced the Ekman spiral $(\tilde{d}_1(0)+i\tilde{d}_2(0))e^{(1+i)\lambda z}$. Instead in our analysis, it is shown that only a fraction of the ice-velocity (namely the velocity of ice with respect of the TDC) acts as a forcing for the ice-drift component of the Ekman spiral $({d}_1(0)+i{d}_2(0))e^{(1+i)\lambda z}$ in
\begin{equation}  
\begin{aligned}
u+iv=(d_1(0)+m\cos(\alpha))e^{(1+i)\lambda z}+i(d_2(0)+m\sin(\alpha))e^{(1+i)\lambda z}+\\
+c e^{k(b+z)}e^{ik(a-z-ct)}+(\mathfrak{u}_g+i\mathfrak{v}_g),
\end{aligned}
\end{equation}
thus implying that the amplitude of the nonlinear ice-drift current at the North Pole is smaller that in other arctic regions (for the same surface stress), due to the presence of the Transpolar Drift Current.\\
Moreover, note that the coordinates used in \cite{Constantin2022} and \cite{Constantin2022NOTE} are the standard Cartesian coordinates for the tangent plane approximation, based on the classical spherical coordinates ($x$ pointing to East and $y$ pointing to North), while our analysis is built upon the use of the ``new" rotated spherical coordinates developed in \cite{CJ2023}, and here reviewed in Appendix \ref{rotated spherical}. In spite of this difference, the qualitative properties of the solution we found in our analysis and the ones in \cite{Constantin2022} match.\\
Finally, our solution \eqref{postion} shows an interesting feature: in \eqref{TDC} we described the Ekman-type part of the Transpolar Drift Current as
\begin{equation}
    me^{i\alpha}e^{(1+i)sz}
\end{equation}
with $s\in\mathbb{R}^+$, but in the nonlinear solution \eqref{postion} and subsequently in \eqref{38} instead of $s$, the exponent features the presence of $\lambda$ (namely the inverse of the Ekman depth), as a result of the pressure gradient being balanced by the geostrophic current in \eqref{governing enqs}, leading to the system of ODEs \eqref{32}. Consequently, if $s\not=\lambda$ we can conclude that only the value of the Ekman-type current of the TDC at the surface influenced the Ekman spiral of the nonlinear ice-drift flow. On the other hand, if the Ekman-type current of the TDC is a ``pure" Ekman spiral, i.e. $s=\lambda$, the entirety of it is present in the solution. This result is a consequence of the fact that $\left(m e^{\lambda z}\cos(\lambda z+\alpha),\  m e^{\lambda z}\sin(\lambda z+\alpha)\right)$ is a solution of the ``classical" system of differential equations for the Ekman spiral (see e.g. \cite{Kundu} or \cite{Pedlosky})
\begin{equation}
	\left\{ \begin{aligned}	
		- f\mathsf{v}= 
        A_V \frac{\partial^2 \mathsf{u}}{\partial z^2},\\
		f\mathsf{u}=
        A_V \frac{\partial^2 \mathsf{v}}{\partial z^2},
	\end{aligned}\right.
\end{equation}
recalling that $\lambda=\sqrt{\frac{f}{2A_V}}$. This, applied to \eqref{governing enqs 2}, leads \eqref{32} to became
\begin{equation}\label{32ter}
	\left\{ \begin{aligned}	
		0&=f d_2(z)+A_V d_1''(z)+\left(fc + 2 A_V k^2 c - k c^2\right)e^{k(b+z)}\sin(\xi),\\
	0&=-f d_1(z)+A_V d_2''(z) -\left(fc + 2 A_V k^2 c - k c^2\right)e^{k(b+z)}\cos(\xi),
	\end{aligned}\right.
\end{equation}
whose solution is $d(z)=(d_1(0)e^{(1+i)\lambda z}, d_2(0)e^{(1+i)\lambda z})$, provided the dispersion relation \eqref{34}. Inserting such a solution into \eqref{Lag} gives exactly the same expression for the solution to \eqref{governing enqs} and \eqref{scaled cont}, namely
\begin{equation}
    \left\{
    \begin{aligned}	
        x(t; a,b,z) &= a + \left[\left(d_1(0)+ m \cos(\alpha)\right)e^{(1+i)\lambda z}+\mathfrak{u}_g\right]t -\\
        &\hspace{3cm}-\frac{1}{k} e ^{k(b+z)}\sin(k(a-z-ct)), \\
        y(t; a,b,z) &= b + \left[\left(d_2(0)+  m \sin(\alpha)\right)e^{(1+i)\lambda z}+\mathfrak{v}_g\right]t  +\\
            &\hspace{3cm} +\frac{1}{k} e ^{k(b+z)}\cos(k(a-z-ct)) .
    \end{aligned}
    \right.
\end{equation}
Therefore we can state that only if $s=\lambda$, the whole Ekman component of the TDC is present in the solution of the nonlinear ice-drift problem.\\
The structure of \eqref{32} also shows that, if a non-geostrophic constant current is present, it is forced to ``transform" into an Ekman spiral, provided that the pressure perturbation $\hat{p}$ or its gradient to be zero. More precisely, assuming
\begin{equation}\label{LagNONGEO}
	\left\{ \begin{aligned}	
		x(t; a,b,z)=a + d_1(z)t -\frac{1}{k} e ^{k(b+z)}\sin(k(a-z-ct)) + \overline{\mathfrak{u}}t\\
			y(t; a,b,z)=b + d_2(z)t +\frac{1}{k} e ^{k(b+z)}\cos(k(a-z-ct)) +\overline{\mathfrak{v}}t
	\end{aligned}\right.
\end{equation}
with $\overline{\mathfrak{u}}, \overline{\mathfrak{v}}$ representing a non-geostrophic constant current. We have omitted the TDC for simplicity. Given the dispersion relation \eqref{34} and repeating the same arguments as in the previous section and assuming again the pressure perturbation $\hat{p}$ or its gradient to be zero, we get the system
\begin{equation}
	\left\{ \begin{aligned}	
		&A_V d_1''(z)+f d_2(z)=-f\overline{\mathfrak{v}},\\
		&A_V d_2''(z)-f d_1(z)=f\overline{\mathfrak{u}},
	\end{aligned}\right.
\end{equation}
which, in complex-variable notation,
\begin{equation}
d(z)=d_1(z) + i d_2(z)\qquad \text{and}\qquad \gamma:=-\frac{f}{A_V}\overline{\mathfrak{v}} + i\frac{f}{A_V}\overline{\mathfrak{u}}
\end{equation}
is equivalent to
\begin{equation}\label{complex NG}
d''(z)-i\frac{f}{A_V} d(z)=\gamma.
\end{equation}
The solution to \eqref{complex NG}, recalling that ice-drift currents are insignificant at great depth, is
\begin{equation}
	d(z)=[d(0)+\overline{\mathfrak{u}}+i\overline{\mathfrak{v}}]e^{(1+i)\lambda z}-\overline{\mathfrak{u}}-i\overline{\mathfrak{v}},
\end{equation}
where $d(0)\in\mathbb{C}$ is the ice-drift current at the surface. Therefore \eqref{LagNONGEO} becomes
\begin{equation}
	\left\{ \begin{aligned}	
		x(t; a,b,z)=a + [d_1(0)+\overline{\mathfrak{u}}]e^{(1+i)\lambda z}t -\frac{1}{k} e ^{k(b+z)}\sin(k(a-z-ct)), \\
			y(t; a,b,z)=b + [d_2(0)+\overline{\mathfrak{v}}]e^{(1+i)\lambda z}t +\frac{1}{k} e ^{k(b+z)}\cos(k(a-z-ct)) ,
	\end{aligned}\right.
\end{equation}
showing that the non-geostrophic current $\overline{\mathfrak{u}}, \overline{\mathfrak{v}}$ is forced to ``transform" into an Ekman spiral.

\section*{Acknowledgments}
The author is grateful to Luigi Roberti for helpful comments and to the referees for the useful suggestions.
\section*{Declarations}

\begin{itemize}
\item Funding: This research was supported by the Austrian Science Fund (FWF) [grant
number Z 387-N].
\item Conflict of interest/Competing interests: The author declares no conflict of interest.
\item Ethics approval and consent to participate: Not applicable
\item Data availability : Not applicable
\item Materials availability : Not applicable
\item Code availability: Not applicable
\item Author contribution: Sole author
\item ORCID: 0009-0008-5454-0922
\end{itemize}

\begin{appendices}
\section{Full derivation of the Navier-Stokes equations}\label{NS-Derivation}
This appendix is devoted to the derivation of the Navier-Stokes equation in spherical coordinates for geophysical flows (atmospheric or oceanic) in its most general setting, correcting the ones in the literature, e.g. in \cite{CJ2023} or \cite{CJChapter}, in order to provide the most complete formulation of the governing equations.\\
The only assumption will be that the vertical and horizontal eddy viscosities \(\mu_V\) and \(\mu_H\) vary with depth but not with respect to the horizontal coordinates, which is physically reasonable since, in most cases, the scales on which the fluid flow is analyzed are much smaller than those on which eddy viscosities vary horizontally. However, extending this to account for viscosities that vary with horizontal coordinates will be almost straightforward, at the price of making the equations more intricate.\\
Before proceeding with the derivation, we briefly recall the formulae for the differential operators in orthogonal curvilinear coordinates used in this paper, namely the gradient and the Laplacian of a scalar field, the divergence of a vector field and of a second order tensor, the gradient of a vector field and the advection operator $\bu\cdot\nabla$. We refer to  \cite{Arfken}, \cite{JOG}, and \cite{Spiegel} for a more complete exposition.
\subsection{Differential Operators in Orthogonal Curvilinear Coordinates}\label{appA}
In this subsection, we will not make the distinction between physical variables (previously denoted with $'$) and non-dimensional variables. For simplicity, everything will be denoted without primes.\\ 
Let us start by setting $(\ea, \eb, \ec)$ a basis of $\mathbb{R}^3$, and Cartesian coordinates $(x, y, z)$. A curvilinear coordinate system $u_1,
 u_2, u_3$ of $\mathbb{R}^3$ is given by the scalar functions
\begin{equation}\label{coordinate change}
	\left\{ \begin{array}{ll}
		u_1=u_1(x,y,z),\\
		u_2=u_2(x,y,z),\\
		u_3=u_3(x,y,z),
	\end{array}\right.\qquad
\left\{ \begin{array}{ll}
x=x(u_1, u_2, u_3),\\
y=y(u_1, u_2, u_3),\\
z=z(u_1, u_2, u_3),\\
\end{array}\right.
\end{equation}
giving the so-called change of coordinates and the inverse change of coordinates. The functions in \eqref{coordinate change}  are assumed to be single-valued and to have continuous derivatives so that there is a one-to-one correspondence between $(x, y, z)$ and $(u_1, u_2, u_3)$.\\
Let $\mathfrak{r}=x\,\ea+y\,\eb+z\,\ec$ be the position vector of a given point $P$, which, due to \eqref{coordinate change} can be written as $\mathfrak{r}=x(u_1, u_2, u_3)\ea+y(u_1, u_2, u_3)\eb+z(u_1, u_2, u_3)\ec$, namely $\mathfrak{r}=\mathfrak{r}(u_1, u_2, u_3)$, 
and let  
\begin{equation}
   \h_i=\left|\frac{\partial\mathfrak{r}}{\partial u_i} \right|,
\end{equation}
denote the scale factors (also known as metric coefficients or Lamé coefficients). The associated unit tangent vectors are given by $\mathfrak{e}_i=\nicefrac{\frac{\partial\mathfrak{r}}{\partial u_i}}{|\frac{\partial\mathfrak{r}}{\partial u_i}|}=\nicefrac{\frac{\partial\mathfrak{r}}{\partial u_i}}{\h_i}$, for $i=1,2,3$.\\
The aforementioned formulae for the differential operators are summarized below.
\begin{itemize}
\item Gradient and Laplacian of a scalar field: given a scalar field $f(u_1,u_2,u_3)$, its gradient is given by
\begin{equation}
	\nabla f=\sum_{i}\frac{1}{\h_i} \frac{\partial f}{\partial u_i}{\mathfrak{e}_i}
\end{equation}
and its Laplacian by
\begin{equation}
	\Delta f= \frac{1}{\h}\left[\frac{\partial }{\partial u_1}\left(\frac{\h_2\h_3 }{ \h_1}\frac{\partial f}{\partial u_1}\right) +\frac{\partial }{\partial u_2}\left(\frac{\h_1\h_3 }{ \h_2}\frac{\partial f}{\partial u_2}\right)  + \frac{\partial }{\partial u_3}\left(\frac{\h_1\h_2 }{ \h_3}\frac{\partial f}{\partial u_3}\right) \right].
\end{equation}
\item Divergence and gradient of a vector field: given a vector field $	V(u_1,u_2,u_3)=\left(V_1(u_1,u_2,u_3), V_2(u_1,u_2,u_3), V_3 (u_1,u_2,u_3)\right)$ with respect to the base $\mathfrak{e}_i$ the divergence of $V$ is a scalar given by
\begin{equation}\label{Div}
	\begin{aligned}
		\nabla\cdot V=\sum_{i=1,2,3} \frac{\h_i^2}{\h}\nabla u_i\cdot \nabla \left(\frac{\h}{\h_i} V_i \right)=\frac{1}{\h}\sum_{i=1,2,3} \frac{\partial}{\partial u_i}\left(\frac{\h}{\h_i} V_i \right),
	\end{aligned}
\end{equation}
while its gradient, which is a second-order tensor (see \cite{JOG}), is given by the relation  $(\nabla V)^T=\nabla\otimes V$ where
\begin{equation}\label{gradVF}
		\nabla \otimes V=\sum_{i=1,2,3}  \frac{\mathfrak{e}_i}{\h_i}\frac{\partial \left(\sum_{j} V_j\mathfrak{e}_j\right)}{\partial u_i}.
\end{equation}
\item Advection operator $\bu\cdot\nabla$: in previous sections, we wrote the velocity vector field $\bu$ as $\bu=(u,v,w)$. However, this notation in the following computations could be confusing, therefore let us write $\bu=(\mathfrak{U}_1, \mathfrak{U}_2, \mathfrak{U}_3)$. Given a vector field $V$, the components of the advection operator $\bu\cdot \nabla$ acting on a vector field $V$ are given by
\begin{equation}\label{C23}
	\begin{aligned}    
		\left\{(\bu\cdot \nabla) V\right\}_i=\sum_{j=1,2,3}\left(\frac{\mathfrak{U_j}}{\h_j}\frac{\partial V_i}{\partial u_j}-\frac{\mathfrak{U_j} V_j}{\h_i\h_j}\frac{\partial \h_j}{\partial u_i}+\frac{\mathfrak{U_i} V_j}{\h_i\h_j}\frac{\partial \h_i}{\partial u_j}\right).
	\end{aligned}
\end{equation}

\item Divergence of a second-order tensor: the formula for the divergence of a second-order tensor $\mathbf{A}=A^{ij}\mathfrak{e}_i \otimes\mathfrak{e}_j$ is given by (see \cite{JOG})
\begin{equation}
	\nabla\cdot \mathbf{A}=\sum_{i,j}\left(\frac{1}{\h_j}\frac{\partial A^{ij}}{\partial u_j}\mathfrak{e}_j + A^{ij}\left(\frac{\omega_j}{\h_j}\times\mathfrak{e}_i\right)+\sum_k A^{ij}\left[\frac{\omega_k}{\h_k}\cdot(\mathfrak{e}_i\times\mathfrak{e}_k)\right]\mathfrak{e}_i\right),
\end{equation}
which, written component by component, reads as
\begin{equation}\label{divT}
	\begin{aligned}
		(\nabla \cdot \mathbf{A})_1 &= \frac{1}{\h_1} \frac{\partial A^{11}}{\partial u_1}+\frac{1}{\h_2} \frac{\partial A^{12}}{\partial u_2}+\frac{1}{\h_3} \frac{\partial A^{13}}{\partial u_3} + \\
		&\quad+\frac{A^{13} + A^{31}}{\h_1 \h_3}\frac{\partial \h_1}{\partial u_3}+\frac{A^{12} + A^{21}}{\h_1\h_2} \frac{\partial \h_1}{\partial u_2}+\frac{A^{13}}{\h_2\h_3} \frac{\partial \h_2}{\partial u_3}\\
		&\quad+ \frac{A^{11} - A^{22}}{\h_1\h_2}\frac{\partial \h_2}{\partial u_1}
		+ \frac{A^{12} }{\h_3\h_2}\frac{\partial \h_3}{\partial u_2}- \frac{A^{33} - A^{11}}{\h_3\h_1}\frac{\partial \h_3}{\partial u_1} \\
		(\nabla \cdot \mathbf{A})_2 &= \frac{1}{\h_1} \frac{\partial A^{21}}{\partial u_1}+\frac{1}{\h_2} \frac{\partial A^{22}}{\partial u_2}+\frac{1}{\h_3} \frac{\partial A^{23}}{\partial u_3} +\\
		&\quad+ \frac{A^{23}}{\h_1 \h_3}\frac{\partial \h_1}{\partial u_3}+\frac{A^{22} - A^{11}}{\h_1\h_2} \frac{\partial \h_1}{\partial u_2}+\frac{A^{23}+A^{32}}{\h_2\h_3} \frac{\partial \h_2}{\partial u_3}+ \\
		&\quad+\frac{A^{12} +A^{21}}{\h_1\h_2}\frac{\partial \h_2}{\partial u_1}
		+ \frac{A^{22} -A^{33}}{\h_3\h_2}\frac{\partial \h_3}{\partial u_2}+\frac{A^{21}}{\h_3\h_1}\frac{\partial \h_3}{\partial u_1} \\
		(\nabla \cdot \mathbf{A})_3 &= \frac{1}{\h_1} \frac{\partial A^{31}}{\partial u_1} +\frac{1}{\h_2} \frac{\partial A^{32}}{\partial u_2}+\frac{1}{\h_3} \frac{\partial A^{33}}{\partial u_3}  + \\
		&\quad+ \frac{A^{33} - A^{11}}{\h_1 \h_3}\frac{\partial \h_1}{\partial u_3}+\frac{A^{32}}{\h_1\h_2} \frac{\partial \h_1}{\partial u_2}+\frac{A^{33}-A^{22}}{\h_2\h_3} \frac{\partial \h_2}{\partial u_3}+ \\
		&\quad+ \frac{A^{31}}{\h_1\h_2}\frac{\partial \h_2}{\partial u_1}
		+ \frac{A^{23}+A^{32} }{\h_3\h_2}\frac{\partial \h_3}{\partial u_2}+ \frac{A^{13} + A^{31}}{\h_3\h_1}\frac{\partial \h_3}{\partial u_1}.
	\end{aligned}
\end{equation}

\end{itemize}

\noindent
Now we turn to the derivation of the equations of motion.\\
Newton's second law, relating the acceleration to the net force at a point, sometimes also called Cauchy's equation of motion in the context of continuum mechanics, in vectorial (coordinate-free) form is given by \cite{Kundu}:
 \begin{equation}\label{NS}
  \rho' \frac{D \mathbf{u}'}{D t'} = \rho' \mathbf{g}' + \nabla' \cdot \boldsymbol{\sigma}' ,
\end{equation}
while the continuity equation is
\begin{equation}\label{c}
  \frac{1}{\rho'} \frac{D \rho'}{D t'} + \nabla' \cdot \mathbf{u'} = 0 ,
\end{equation}
\noindent
where $t'$ is the time, $\rho'$ is the fluid density, $\frac{D}{Dt'}=\frac{\partial}{\partial t' }+\mathbf{u}'\cdot \nabla'$ is the advective derivative, $\mathbf{g}'$ is the gravity acceleration (its modulus being approximately $9.81\, m\, s^{-2}$ at the surface of the Earth) and $\boldsymbol{\sigma}'$ is the stress tensor (a second-order tensor with nine components that describes the internal forces within a solid body under external loads. It comprises three normal stresses and six shear stresses). Recall that we use primes to represent dimensional (physical) variables.\\
In the standard Cartesian coordinates system $(x, y, z)$  (with respect to the basis  $(\mathbf{e_1}, \mathbf{e_2}, \mathbf{e_3}) $ of $\mathbb{R}^3$), the stress tensor's components, for $i,j=1,2,3$, are given by the constitutive equations:
\begin{equation}\label{tau}
    \boldsymbol{\sigma}'_{ij} = - \left(p' + \frac{2}{3} \mu' \nabla' \cdot \mathbf{u}'\right)\delta_{ij} + 2 \mu' \boldsymbol{\epsilon}'_{ij},
\end{equation}
with $\delta_{ij}$ being the Kronecker delta and $\boldsymbol{\epsilon}'_{ij}=\frac{1}{2} \left(\frac{\partial u'_i}{\partial x'_j} + \frac{\partial u'_j}{\partial x'_i}\right)$ the rate of strain tensor, $\mu'$ the dynamic viscosity and $p'$  the pressure . For simplicity of notation the decomposition of $\mu'$ into its vertical and horizontal components will be done at the end. Define
\begin{equation}\label{s}
    \s'=\mu' \left(\nabla' \mathbf{u}' +(\nabla' \mathbf{u}' )^T\right)=2\mu'\boldsymbol{\epsilon}',
\end{equation}
and using \eqref{tau}, \eqref{s} and \eqref{c}, equation \eqref{NS} becomes
\begin{equation}\label{ns1}
\begin{aligned}
 \rho' \left(\frac{\partial}{\partial t' }+\mathbf{u}'\cdot \nabla'\right)\mathbf{u}'&=\rho' \mathbf{g}' -\nabla' p' -\frac{2}{3}\nabla'\left[\mu' \left( \nabla' \cdot \mathbf{u}' \right) 
 \right]+   \nabla' \cdot \left(2\mu' \boldsymbol{\epsilon}' \right)\\
 &=\rho' \mathbf{g}' -\nabla' p' +\frac{2}{3}\nabla'\left[\mu' \left( \frac{1}{\rho'}\frac{D\rho'}{Dt'}\right) 
 \right]+   \nabla' \cdot \mathbb{S}'.\\
 \end{aligned}
\end{equation}
Equation \eqref{ns1} is the general form of the Navier-Stokes equation.\\
It is common to assume $\mu'$ to be constant, hence obtaining 
\begin{equation}\label{kunduNS}
    \rho'\frac{D\bu'}{Dt'}=\rho' \mathbf{g}' -\nabla' p' +\mu'\Delta'\bu' +\frac{1}{3}\mu' \nabla'\left(\nabla'\cdot \bu'\right)
\end{equation} 
which becomes 
\begin{equation}
    \rho'\frac{D\bu'}{Dt'}=\rho' \mathbf{g}' -\nabla' p' +\mu'\Delta'\bu' 
\end{equation} 
if the fluid is incompressible, namely $\nabla'\cdot \bu'=0,$ with $\Delta'$ being the Laplacian.
\subsection {Classical Spherical Coordinates}\label{classical spherical coordinates}
\begin{figure}
    \centering
    \includegraphics[width=0.5\linewidth]{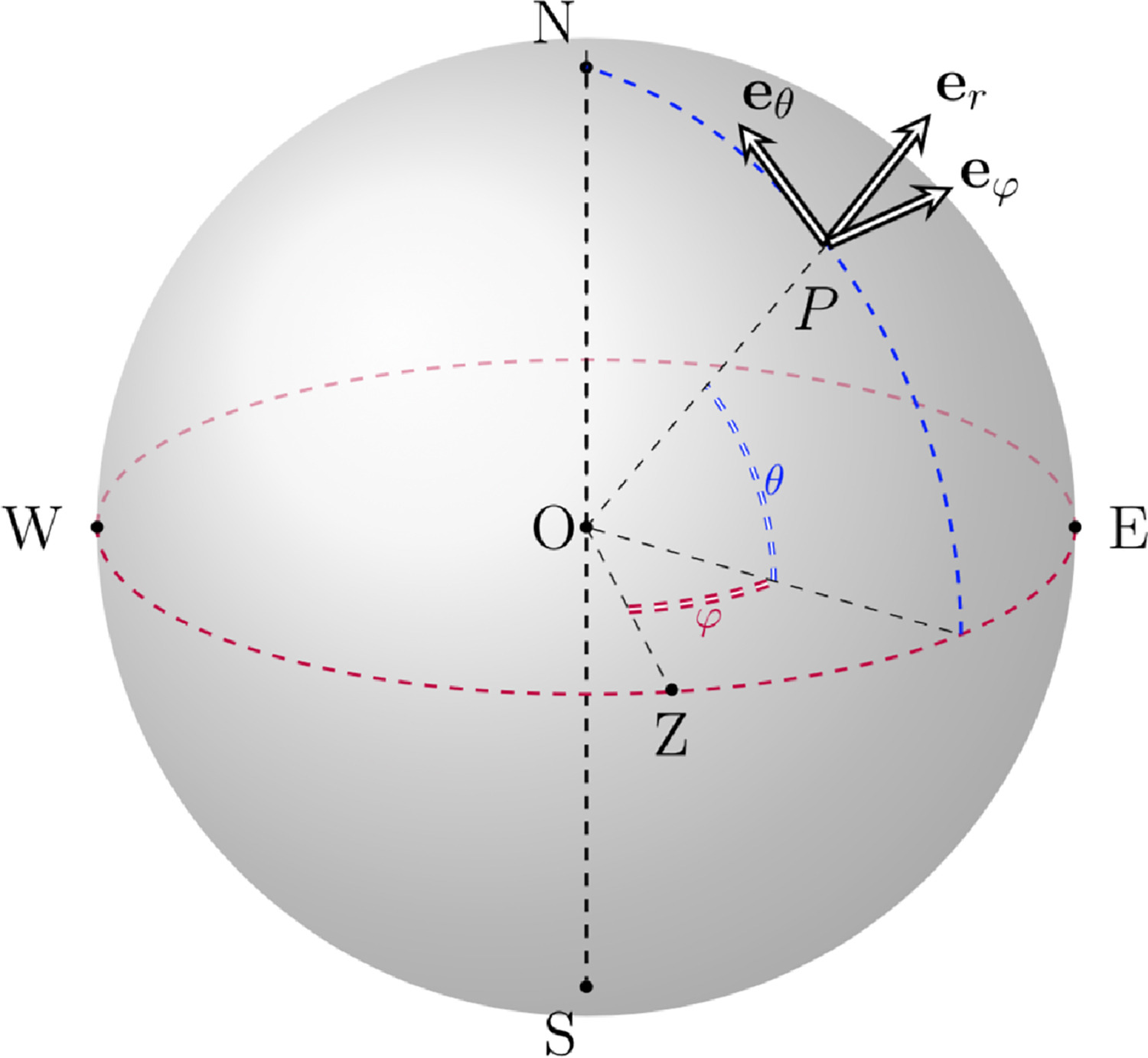}
    \caption{Classical spherical coordinate system. Image from \cite{CJ2023}. CC BY 4.0 (\url{http://creativecommons.org/licenses/by/4.0})}
    \label{fig-spherical}
\end{figure}

A  coordinate system suitable for the Earth, with the poles 
excluded, is the system of (right-handed) spherical coordinates $(\varphi, \theta, r')$: $r'$ is the distance from the Earth’s center, $\theta \in [-\nicefrac{\pi}{2},\nicefrac{\pi}{2}]$ is the angle of latitude and $\varphi \in [0,2\pi]$ is the angle of longitude. The unit vectors in this system
are $(\ep, \et, \er)$ and the corresponding velocity components are $\bu'=(u',v',w')$: $\ep $ points from West to East, $\et $ from South to North and $\er $  points upwards. See Figure \ref{fig-spherical}.
The $(\varphi, \theta, r')$-system is associated with a point fixed on the sphere (other than at the 
two poles where the unit vectors are not well-defined) which is rotating about its polar axis (with
constant angular speed $\Omega' \approx 7.29\cdot10^{-5} \,s^{-1}$).\\
The change of coordinates from spherical to Cartesian and vice-versa is given by

\begin{equation}
\left\{ \begin{array}{ll}
 x'=r' \cos\theta \cos \varphi,\\
      y'=r' \cos\theta \sin \varphi,\\
        z'=r' \sin\theta ,
 \end{array}\right.
\qquad
\left\{ \begin{array}{ll}
  \varphi = \tan^{-1}\left(\frac{y'}{x'}\right),\\
    \theta = \sin^{-1}\left(\frac{z'}{\sqrt{x'^2+y'^2+z'^2}}\right),\\
   r'=\sqrt{x'^2+y'^2+z'^2}.
 \end{array}\right.
  \label{rot coord}
\end{equation}
\noindent

\noindent
and the unit tangent vectors are
\begin{equation}
    \mathbf{e_{\varphi}}=\frac{\frac{\partial\mathfrak{r}}{\partial\varphi}}{\left|\frac{\partial\mathfrak{r}}{\partial\varphi}\right|},\qquad  \mathbf{e_{\theta}}=\frac{\frac{\partial\mathfrak{r}}{\partial\theta}}{\left|\frac{\partial\mathfrak{r}}{\partial\theta}\right|},\qquad  \mathbf{e}_r=\frac{\frac{\partial\mathfrak{r}}{\partial r}}{\left|\frac{\partial\mathfrak{r}}{\partial r}\right|},
\end{equation}
where $\mathfrak{r}=x(\varphi,\theta, r) \ea+y(\varphi,\theta, r) \eb+z(\varphi,\theta, r) \ec$, 
leading to \begin{equation}
\left\{ \begin{array}{ll}
  \mathbf{e_{\varphi}}=-\sin\varphi \mathbf{e_1}+\cos\varphi \mathbf{e_2},\\
\mathbf{e_{\theta}}=-\cos\varphi\sin\theta \mathbf{e_1}-\sin\varphi\sin\theta \mathbf{e_2}+\cos\theta \mathbf{e_3},\\
    \mathbf{e}_{r}=\cos\varphi\cos\theta \mathbf{e_1}+\sin\varphi\cos\theta \mathbf{e_2}+\sin\theta \mathbf{e_3},
 \end{array}\right.
  \label{base sferiche}
\end{equation}
and
\begin{equation}\label{hh}
\begin{aligned}
& \h_{\varphi}=\left|\frac{\partial\mathfrak{r}}{\partial\varphi}\right|=r' \cos\theta,\qquad
  \h_{\theta}=\left|\frac{\partial\mathfrak{r}}{\partial\theta}\right|=r' ,\qquad
   \h_{r'}=\left|\frac{\partial\mathfrak{r}}{\partial r}\right|=1,\\
&  \h:=\h_{\varphi}\h_{\theta}\h_{r'}=r'^2\cos\theta.
\end{aligned}
\end{equation}
\subsubsection{The momentum equation in classical spherical coordinates}
\noindent
Starting from \eqref{ns1}:
\begin{equation}
 \rho' \left(\frac{\partial}{\partial t' }+\mathbf{u}'\cdot \nabla'\right)\mathbf{u}'=\rho' \mathbf{g}' -\nabla' p' +\frac{2}{3}\nabla'\left[\mu' \left( \frac{1}{\rho'}\frac{D\rho'}{Dt'}\right) 
 \right]+   \nabla' \cdot \mathbb{S}'
\end{equation}
we now compute the various terms in spherical coordinates.\\
The gradient in spherical coordinates with respect to the basis $(\mathbf{e_{\varphi}},  \mathbf{e_{\theta}},\mathbf{e_{r}})$ reads
\begin{equation}
   \nabla'= \left( \frac{1}{r'\cos\theta}\frac{\partial}{\partial\varphi},\  \frac{1}{r'}\frac{\partial}{\partial\theta} ,\   \frac{\partial}{\partial r'}
  \right),
\end{equation}
while the advection term is (see Appendix \ref{appA})
\begin{equation}\label{unablau}
    (\bu'\cdot\nabla')\bu'=\left(\frac{u'}{r'\cos\theta}\frac{\partial}{\partial\varphi}+  \frac{v'}{r'}\frac{\partial}{\partial\theta} +w'\frac{\partial}{\partial r'}\right)\begin{pmatrix}
        u'\\ v'\\ w'
    \end{pmatrix} +\frac{1}{r'}\begin{pmatrix}-u'v'\tan \theta + u'w' \\
u'^2\tan\theta+v'w'\\
-u'^2-v'^2
\end{pmatrix}.
\end{equation}
The last term in the previous equation \eqref{unablau} is called metric term and represents curvature effects of the Earth (see \cite{DynamicMeteo}).\\
The gradient of a vector is a tensor and can be computed as (see equation \eqref{gradVF} in Appendix \ref{appA} ):
\begin{equation}\label{nablabu}
    \begin{aligned}
       (\nabla' \bu')^T = &\nabla' \otimes \bu'= \left( \mathbf{e_{\varphi}} \frac{1}{r'\cos\theta}\frac{\partial}{\partial\varphi}+ \mathbf{e_{\theta}} \frac{1}{r'}\frac{\partial}{\partial\theta}+\mathbf{e_{r}}\frac{\partial}{\partial r'}\right)\otimes \\
       &\otimes\left( u'\mathbf{e_{\varphi}} + v'\mathbf{e_{\theta}}+w'\mathbf{e_{r}} \right)=\\
       =& \frac{1}{r'\cos\theta}\frac{\partial u'}{\partial\varphi} \ep \otimes \ep   + \frac{u'}{r'\cos\theta}  \ep\otimes(\sin\theta \et-\cos\theta \er) + \\
       &+ \frac{1}{r'\cos\theta}\frac{\partial v'}{\partial\varphi}   \ep \otimes \et + \frac{v'}{r'\cos\theta}\ep\otimes(-\sin\theta \ep) + \\
       & +\frac{1}{r'\cos\theta}\frac{\partial w'}{\partial\varphi} \ep \otimes \er   + \frac{w'}{r'\cos\theta}  \ep\otimes(\cos\theta \ep) \\
    &+ \frac{1}{r'}\frac{\partial u'}{\partial\theta}  \et \otimes \ep  +\frac{1}{r'}\frac{\partial v'}{\partial\theta}  \et \otimes \et +\frac{v'}{r'}\et\otimes(-\er)   + \\
    &+\frac{1}{r'}\frac{\partial w'}{\partial\theta} \et \otimes \er  + \frac{w'}{r'}\et\otimes\et+ \frac{\partial u'}{\partial r'}\er \otimes\ep  +\\
    &+ \frac{\partial v'}{\partial r'}  \er \otimes \et  + \frac{\partial w' }{\partial r'} \er \otimes \er      .
    \end{aligned}
\end{equation}
Since $\mathbb{S}'=\mu' (\nabla' \bu'+(\nabla' \bu')^T)$ is a symmetric tensor, the component of  $\mathbb{S}'$ 
are\footnote{Note that these components are contravariant due to how they are constructed in \eqref{nablabu}. However, for the sake of notation, we write lower indices instead of upper ones.}:
\begin{equation}    
\begin{aligned}
\s'_{\varphi\varphi}=&2\mu'\left( \frac{1}{r'\cos\theta}\frac{\partial u'}{\partial\varphi} -\frac{v'}{r'}\tan\theta +\frac{w'}{r'}\right),\quad
\s'_{\theta\theta}=2\mu'\left( \frac{1}{r'}\frac{\partial v'}{\partial\theta} + \frac{w'}{r'}\right),\\
 \s'_{rr}=&2\mu'\left(  \frac{\partial w' }{\partial r'} \right),\quad
\s'_{\varphi\theta}=\s'_{\theta\varphi }= \mu'\left(\frac{1}{r'}\frac{\partial u'}{\partial\theta} +\frac{u'}{r'}\tan\theta +  \frac{1}{r'\cos\theta}\frac{\partial v'}{\partial\varphi}\right), \\
\s'_{\varphi r}=&\s'_{r\varphi}=\mu'\left(-\frac{u'}{r'} +\frac{1}{r'\cos\theta}\frac{\partial w'}{\partial\varphi}+\frac{\partial u'}{\partial r'}\right),\\
\s'_{\theta r}=&\s'_{r\theta} = \mu'\left(-\frac{v'}{r'}  +\frac{1}{r'}\frac{\partial w'}{\partial\theta} + \frac{\partial v'}{\partial r'}\right). \\
   \end{aligned}
   \end{equation}
Note that $\nabla'\cdot\s'$ is a vector, and is computed applying \eqref{divT} to $\s'$ and using \eqref{hh}.\\
\noindent
Splitting into the three components we get:\\
$\ep$-component:
\begin{equation}\label{ep}
\begin{aligned}
&\frac{1}{r'^2\cos\theta}\left[\frac{\partial}{\partial\varphi}\left(r'\s'_{\varphi\varphi}\right) + \frac{\partial}{\partial\theta}\left(r'\cos\theta\s'_{\theta\varphi}\right) + \frac{\partial}{\partial r'}\left(r'^2\cos\theta\s'_{r\varphi}\right)\right]+\\
&\quad+\frac{\s'_{\varphi\theta}(-r'\sin\theta)}{r'^2\cos\theta}+\frac{\s'_{r\varphi}\cos\theta}{r'\cos\theta}=\\
&=\frac{1}{r'\cos\theta}\frac{\partial\s'_{\varphi\varphi}}{\partial\varphi} + \frac{1}{r'\cos\theta}\frac{\partial}{\partial\theta}{(\cos\theta\s'_{\theta\varphi})} + \frac{1}{r'^2}\frac{\partial}{\partial r'}{(r'^2\s'_{r\varphi})} +\\
&\quad +\frac{\s'_{r\varphi}-\s'_{\varphi\theta}\tan\theta}{r'}
=\\
&=\underbrace{\frac{\mu'}{r'^2\cos^2\theta}\frac{\partial^2 u'}{\partial\varphi^2} -\frac{\mu' }{r'^2}\tan\theta\frac{\partial u'}{\partial\theta} +\frac{\mu' }{r'^2}\frac{\partial^2 u'}{\partial\theta^2}+2\frac{\mu'}{r'}\frac{\partial u}{\partial r'}+\mu'\frac{\partial^2 u'}{\partial r'^2}}_{=\Laplace'_{\mu}u'}+\\
&\quad+\left[ -\frac{2\mu'}{r'^2}\frac{\tan\theta }{\cos\theta}\frac{\partial v'}{\partial\varphi}+\frac{2\mu'}{r'^2\cos\theta}\frac{\partial w'}{\partial\varphi} -\frac{\mu'}{r'^2}u'-\frac{\mu' \tan^2\theta}{r'^2}u'\right]+\\
&\quad+\left[\frac{d\mu'}{d r'}\left( \frac{\partial u'}{\partial r'}-\frac{u'}{r'}+\frac{1}{r'\cos\theta}\frac{\partial w'}{\partial\varphi}\right)\right] +\frac{\mu'}{r'^2\cos^2\theta}\frac{\partial^2 u'}{\partial\varphi^2}+ \frac{\mu'}{r'^2\cos\theta}\frac{\partial^2 v'}{\partial\theta\partial\varphi}+\\
&\quad+\frac{\mu'}{r'^2\cos\theta}\frac{\partial w'}{\partial\varphi}+\frac{\mu'}{r'\cos\theta}\frac{\partial^2 w'}{\partial r' \partial\varphi}+\frac{\mu'}{r'^2\cos\theta}\frac{\partial w'}{\partial\varphi}- \frac{\mu' \tan\theta}{r'^2\cos\theta}\frac{\partial v'}{\partial\varphi},
\end{aligned}
\end{equation}
$\et$-component:
\begin{equation}\label{et}
\begin{aligned}
&\frac{1}{r'^2\cos\theta}\left[\frac{\partial}{\partial\varphi}\left(r'\s'_{\theta\varphi}\right) + \frac{\partial}{\partial\theta}\left(r'\cos\theta\s'_{\theta\theta}\right) + \frac{\partial}{\partial r'}\left(r'^2\cos\theta\s'_{\theta r}\right)\right]+\\
&\quad+\frac{\s'_{\theta r}}{r'}+\frac{\s'_{\varphi\varphi}r'\sin\theta}{r'^2\cos\theta}=\\
&=\frac{1}{r'\cos\theta}\frac{\partial\s'_{\theta\varphi}}{\partial\varphi} + \frac{1}{r'\cos\theta}\frac{\partial}{\partial\theta} {(\cos\theta\s'_{\theta\theta})}+ \frac{1}{r'^2}\frac{\partial}{\partial r'}{(r'^2\s'_{\theta r})} +\\
&\quad+\frac{\s'_{\theta r}+\s'_{\varphi\varphi}\tan\theta}{r'}
=\\
&=\underbrace{\frac{\mu'}{r'^2\cos^2\theta}\frac{\partial^2 v'}{\partial\varphi^2} -\frac{\mu' }{r'^2}\tan\theta\frac{\partial v'}{\partial\theta} +\frac{\mu' }{r'^2}\frac{\partial^2 v'}{\partial\theta^2}+2\frac{\mu'}{r'}\frac{\partial v'}{\partial r'}+\mu'\frac{\partial^2 v'}{\partial r'^2}}_{=\Laplace'_{\mu}v'} +\\
& \quad+\left[\frac{2\mu'}{r'^2}\frac{\partial w'}{\partial\theta} -\frac{\mu' v'}{r'^2}+\frac{2\mu'\tan\theta}{r'^2\cos\theta}\frac{\partial u'}{\partial \varphi}-\frac{\mu'\tan^2\theta}{r'^2}v'\right]+\\
&\quad+\left[\frac{d\mu'}{dr'}\left(\frac{\partial v'}{\partial r'}-\frac{v'}{r'}+\frac{1}{r'}\frac{\partial w'}{\partial \theta}\right)\right]+\frac{\mu'\tan\theta}{r'^2\cos\theta}\frac{\partial u'}{\partial \varphi}+\frac{\mu'}{r'^2\cos\theta}\frac{\partial^2 u'}{\partial \varphi\partial\theta} +\\
&\quad+ \frac{\mu'}{r'^2}\frac{\partial^2v'}{\partial\theta^2}-\frac{\mu'}{r'^2}\tan\theta\frac{\partial v'}{\partial\theta}-\frac{\mu'}{r'^2}v'+2\frac{\mu'}{r'^2}\frac{\partial w'}{\partial \theta}+\frac{\mu'}{r'}\frac{\partial^2 w' }{\partial r' \partial\theta}-\frac{\mu'\tan^2\theta}{r'^2}v',
\end{aligned}
\end{equation}
$\er$-component:
\begin{equation}\label{er}
\begin{aligned}
&\frac{1}{r'^2\cos\theta}\left[\frac{\partial}{\partial\varphi}\left(r'\s'_{r\varphi}\right) + \frac{\partial}{\partial\theta}\left(r'\cos\theta\s'_{r\varphi}\right) + \frac{\partial}{\partial r'}\left(r'^2\cos\theta\s'_{rr}\right)\right]-\\
&\quad-\frac{\s'_{\varphi\varphi}\cos\theta}{r'\cos\theta}-\frac{\s'_{\theta\theta}}{r'}=\\
&=\frac{1}{r'\cos\theta}\frac{\partial\s'_{r\varphi}}{\partial\varphi} + \frac{1}{r'\cos\theta}\frac{\partial}{\partial\theta}{(\cos\theta\s'_{r\theta})} + \frac{1}{r'^2}\frac{\partial}{\partial r'}{(r'^2\s'_{rr})} -\\
&\quad -\frac{\s'_{\varphi\varphi}+\s'_{\theta\theta}}{r'}
=\\
&=\underbrace{\frac{\mu'}{r'^2\cos^2\theta}\frac{\partial^2 w'}{\partial\varphi^2} -\frac{\mu' }{r'^2}\tan\theta\frac{\partial w'}{\partial\theta} +\frac{\mu' }{r'^2}\frac{\partial^2 w'}{\partial\theta^2}+2\frac{\mu'}{r'}\frac{\partial w'}{\partial r'}+\mu'\frac{\partial^2 w'}{\partial r'^2}}_{=\Laplace'_{\mu}w'} +\\
&\quad+\left[-\frac{2\mu'}{r'^2\cos\theta}\frac{\partial u'}{\partial\varphi} +\frac{2\mu'}{r'^2}\tan\theta v' -2\frac{\mu'}{r'^2}w'-\frac{2\mu'}{r'^2}\frac{\partial v'}{\partial\theta}\right]+\left[2\frac{d\mu'}{dr'}\frac{\partial w'}{\partial r'}\right]-\\
&\quad-\frac{\mu'}{r'^2\cos\theta}\frac{\partial u'}{\partial \varphi}+\frac{\mu'}{r'\cos\theta}\frac{\partial^2 u'}{\partial \varphi\partial r'}+\frac{\mu'}{r'^2}\tan\theta v' -\frac{\mu'}{r'}\tan\theta \frac{\partial v'}{\partial r'}-\frac{\mu'}{r'^2} \frac{\partial v'}{\partial\theta} +\\
&\quad+\frac{\mu'}{r'} \frac{\partial^2 v'}{\partial\theta\partial r'}+\frac{2\mu'}{r'}\frac{\partial w'}{\partial r'}  +\mu'\frac{\partial^2 w'}{\partial r'^2} -2\frac{\mu'}{r'^2}w',
\end{aligned}
\end{equation}
where  
\begin{equation}
\Laplace'_{\mu}:= \left[\mu' \left(\frac{\partial^2}{\partial r'^2}+ \frac{2}{r'}\frac{\partial}{\partial r'}\right) + \frac{\mu'}{r'^2}\left(\frac{1}{ \cos^2 \theta} \frac{\partial^2 }{\partial \varphi^2}+\frac{\partial^2}{\partial \theta^2}   -  \tan\theta\frac{\partial}{\partial \theta}\right) \right].
\end{equation}
Now, observe that, computing $\nabla'(\mu' \nabla'\cdot\bu')$ and recalling that we are considering $\mu'=\mu'(r)$, it follows:
\begin{equation}\label{NSlastterms1}
    \begin{aligned}
          &\frac{1}{r'\cos\theta}\frac{\partial}{\partial\varphi}\left[\mu'\left( \frac{1}{r' \cos \theta} \frac{\partial u'}{\partial \varphi} + \frac{1}{r' \cos \theta} \frac{\partial}{\partial \theta} \left( v' \cos \theta  \right) + \frac{1}{r'^2} \frac{\partial}{\partial r'} \left( r'^2 w' \right)\right)\right]=\\
    &=\frac{\mu'}{r'^2 \cos ^2\theta} \frac{\partial^2 u'}{\partial \varphi^2}+\frac{\mu'}{r'^2 \cos \theta} \frac{\partial^2 v'}{\partial \varphi\partial\theta} - \frac{\mu'}{r'^2 \cos \theta} \tan\theta\frac{\partial v'}{\partial \varphi} +\\
    &\quad+\frac{\mu'}{r'\cos\theta}\frac{\partial^2 w'}{\partial\varphi\partial r'}+ \frac{2\mu'}{r'^2\cos\theta} \frac{\partial w'}{\partial \varphi},
\end{aligned}
\end{equation}
    \begin{equation}\label{NSlastterms2}
    \begin{aligned}
   & \frac{1}{r'}\frac{\partial}{\partial\theta}\left[\mu'\left( \frac{1}{r' \cos \theta} \frac{\partial u'}{\partial \varphi} + \frac{1}{r' \cos \theta} \frac{\partial}{\partial \theta} \left( v' \cos \theta  \right) + \frac{1}{r'^2} \frac{\partial}{\partial r'} \left( r'^2 w' \right)\right)\right]=\\
  & =\frac{\mu'\tan\theta}{r'^2\cos\theta}\frac{\partial u'}{\partial \varphi}+\frac{\mu'}{r'^2\cos\theta}\frac{\partial^2 u'}{\partial \varphi\partial\theta} + \frac{\mu'}{r'^2}\frac{\partial^2v'}{\partial\theta^2}-\frac{\mu'}{r'^2}\tan\theta\frac{\partial v'}{\partial\theta}-\\
  &\quad -\frac{\mu'}{r'^2\cos^2\theta}v'+2\frac{\mu'}{r'^2}\frac{\partial w'}{\partial \theta}+\frac{\mu'}{r'}\frac{\partial^2 w' }{\partial r' \partial\theta},
\end{aligned}
\end{equation}
  \begin{equation}\label{NSlastterms3}
    \begin{aligned}
 & \frac{\partial}{\partial r'}\left[\mu'\left( \frac{1}{r' \cos \theta} \frac{\partial u'}{\partial \varphi} + \frac{1}{r' \cos \theta} \frac{\partial}{\partial \theta} \left( v' \cos \theta  \right) + \frac{1}{r'^2} \frac{\partial}{\partial r'} \left( r'^2 w' \right)\right)\right]=\\
 &=\left(\frac{d\mu'}{d r'}\right)\left( \frac{1}{r' \cos \theta} \frac{\partial u'}{\partial \varphi} + \frac{1}{r' \cos \theta} \frac{\partial}{\partial \theta} \left( v' \cos \theta  \right) + \frac{1}{r'^2} \frac{\partial}{\partial r'} \left( r'^2 w' \right)\right)+\\
 &\quad+\mu'\frac{\partial}{\partial r'}\left[\left( \frac{1}{r' \cos \theta} \frac{\partial u'}{\partial \varphi} + \frac{1}{r' \cos \theta} \frac{\partial}{\partial \theta} \left( v' \cos \theta  \right) + \frac{1}{r'^2} \frac{\partial}{\partial r'} \left( r'^2 w' \right)\right)\right]=\\
 &=\left(\frac{d\mu'}{d r'}\right)\nabla'\cdot\bu'-\frac{\mu'}{r'^2\cos\theta}\frac{\partial u'}{\partial \varphi}+\frac{\mu'}{r'\cos\theta}\frac{\partial^2 u'}{\partial \varphi\partial r'}+\mu'\frac{\partial^2 w'}{\partial r'^2} +\frac{2\mu'}{r'}\frac{\partial w'}{\partial r'} -\\
 &\quad-2\frac{\mu'}{r'^2}w'+\frac{\mu'}{r'^2}\tan\theta v'-\frac{\mu'}{r'}\tan\theta \frac{\partial v'}{\partial r'}-\frac{\mu'}{r'^2} \frac{\partial v'}{\partial\theta} +\frac{\mu'}{r'} \frac{\partial^2 v'}{\partial\theta\partial r'}.
    \end{aligned}
\end{equation}
Comparing \eqref{NSlastterms1}, \eqref{NSlastterms2}, \eqref{NSlastterms3} with the last terms of \eqref{ep}, \eqref{et}, \eqref{er}, using the continuity equation \eqref{c} to write $  \mu' \nabla' \cdot \mathbf{u}' = -\frac{\mu'}{\rho'} \frac{D \rho'}{D t'}$, and recalling that the divergence of the vector field $\bu'=(u',v',w')$ in spherical coordinates $(\varphi,\theta, r')$ is
\begin{equation}
	\begin{aligned}	
	\nabla'\cdot\bu'&=\frac{1}{r' \cos \theta} \frac{\partial u'}{\partial \varphi} + \frac{1}{r' \cos \theta} \frac{\partial}{\partial \theta} \left( v' \cos \theta \right) + \frac{1}{r'^2} \frac{\partial}{\partial r'} \left( r'^2 w'\right) =\\
	&= \frac{1}{r' \cos \theta} \frac{\partial u'}{\partial \varphi} + \frac{1}{r'} \frac{\partial v'}{\partial \theta} - \frac{v'}{r'} \tan \theta + \frac{\partial w'}{\partial r'} + \frac{2}{r'} w',
\end{aligned}
\end{equation}
the momentum equation in spherical coordinates becomes
\begin{equation}\label{NS corrected 1}
\begin{aligned}
&\quad \rho'\frac{D}{Dt'}\begin{pmatrix}
u' \\
v' \\
w'\\
\end{pmatrix} +\frac{\rho'}{r'}\begin{pmatrix}-u'v'\tan \theta + u'w' \\
u'^2\tan\theta+v'w'\\
-u'^2-v'^2
\end{pmatrix}+\\
&\quad+2\rho'\Omega'\begin{pmatrix}
    -v'\sin\theta+w'\cos\theta\\
    u'\sin\theta\\
    -u'\cos\theta
\end{pmatrix}+\rho' r' \Omega'^2\begin{pmatrix}
    0\\
    \sin\theta\cos\theta\\
    -\cos^2\theta
\end{pmatrix} =\\
&=-\nabla' p' +\rho'\begin{pmatrix}
    0\\
    0\\
     -g'\frac{R'^2}{ r'^2}\\
\end{pmatrix}+
\Delta'_{\mu}
\begin{pmatrix}
    u'\\ v'\\ w'
\end{pmatrix}-\frac{1}{3}\begin{pmatrix}
    \frac{1}{r'\cos \theta}\frac{\partial}{\partial \varphi}\left(\frac{\mu'_H}{\rho'} \frac{D\rho'}{D t'}\right)\\
    \frac{1}{r'}\frac{\partial}{\partial \theta}\left(\frac{\mu'_H}{\rho'} \frac{D\rho'}{D t'}\right)\\
    \frac{\partial}{\partial r'}\left(\frac{\mu'_V}{\rho'} \frac{D\rho'}{D t'}\right)
\end{pmatrix} \\
&\quad-\frac{1}{r'^2\cos^2\theta}\begin{pmatrix}
    \mu'_H u\\
    \mu'_H v\\
    2\mu'_V(w'\cos^2\theta-v'\sin\theta\cos\theta)
\end{pmatrix}+\frac{2\mu'_H}{r'^2}\frac{\partial}{\partial \theta}\begin{pmatrix}
0\\
 w'\\
   -v'
\end{pmatrix}\\
&\quad+\frac{2\mu'_H}{r'^2\cos\theta}\frac{\partial}{\partial \phi}\begin{pmatrix}
w'-v'\tan\theta\\
u'\tan\theta\\
   -u'
\end{pmatrix}+ \frac{d \mu'_V}{dr'}r' \begin{pmatrix}
    \frac{\partial}{\partial r'}\left(\frac{u'}{r'}\right)\\
       \frac{\partial}{\partial r'}\left(\frac{v'}{r'}\right)\\
        {0}
\end{pmatrix}+\\
&\quad+\frac{d \mu'_H}{d r'} \begin{pmatrix}
    \frac{1}{r'\cos \theta}\frac{\partial  {w'}}{\partial \phi}\\
     \frac{1}{r'} \frac{\partial {w'}}{\partial \theta}\\
{0}
\end{pmatrix}+ \frac{d \mu'_V}{dr'} \begin{pmatrix}
   0\\
   0\\
   2 \frac{\partial w'}{\partial r'}  
\end{pmatrix}+\\
&\quad+\frac{d \mu'_V}{dr'}\begin{pmatrix}
   0\\
   0\\
   \frac{1}{r' \cos \theta} \frac{\partial u'}{\partial \varphi} + \frac{1}{r' \cos \theta} \frac{\partial}{\partial \theta} \left( v' \cos \theta  \right) + \frac{1}{r'^2} \frac{\partial}{\partial r'} \left( r'^2 w' \right)
\end{pmatrix},
\end{aligned}
\end{equation}
where 
\begin{equation}
    \begin{aligned}
      \nabla'&= \left( \frac{1}{r'\cos\theta}\frac{\partial}{\partial\varphi},\  \frac{1}{r'}\frac{\partial}{\partial\theta} ,\   \frac{\partial}{\partial r'}
      \right),\\
        \frac{D}{Dt'}&=\frac{\partial} {\partial t'}+ \frac{u'}{r' \cos \theta} \frac{\partial }{\partial \varphi} + \frac{v'}{r'} \frac{\partial }{\partial \theta} + w'\frac{\partial }{\partial r'}, \\
        \Delta'_{\mu}&=\mu'_V \left(\frac{\partial^2}{\partial r'^2}+ \frac{2}{r'}\frac{\partial}{\partial r'}\right) + \frac{\mu'_H}{r'^2}\left(\frac{1}{ \cos^2 \theta} \frac{\partial^2 }{\partial \varphi^2}+\frac{\partial^2}{\partial \theta^2}   -  \tan\theta\frac{\partial}{\partial \theta}\right),
    \end{aligned}
\end{equation}
$R' \approx 6371\ km$ is the Earth's radius, $g' \approx 9.81 \ m\, s^{-2}$ is the average acceleration of gravity at the surface of the Earth, and we have now made the distinction between $\mu'_V$ (vertical) and $\mu'_H$ (horizontal).\\
The quantities $\mu'_V$ and $\mu'_H$ are the so-called dynamics (effective) viscosities, the sum of the molecular and eddy (related to Reynolds stresses) viscosities \cite{Pedlosky}. The eddy viscosity is several orders of magnitude bigger than the molecular viscosity and is a model for the so-called closure problem in turbulence modeling. Eddy viscosity models are based on two key principles: the Boussinesq hypothesis (stating that turbulent fluctuations dissipate mean flow energy), and the eddy viscosity hypothesis (stating that Reynolds (shear) stresses are proportional to the mean velocity gradient). See \cite{Jiang2020} for a  summary of  recent advances in the mathematical foundations of eddy viscosity models of turbulence.\\
For oceans, the kinematic eddy viscosity  $A'_V:=\frac{\mu'_V}{\rho'}$ varies between $10^{-4}\ m^2\, s^{-1}$ and $10^{-1}\ m^2\, s^{-1}$, while $A'_H:=\frac{\mu'_H}{\rho'}$ ranges between $10\ m^2\, s^{-1}$ and $10^{4}\ m^2\, s^{-1}$.\\ 
In case of atmospheric flows, values of $A'_V$ are $10\ m^2\, s^{-1}$ near the Earth's surface are considerably lower in the atmosphere, while $A'_H$ has order of $10\ m^2\, s^{-1}$ \cite{Pedlosky}.\\
As the eddy viscosity reflects ``the intensity of the whirling agitation" (see \cite{16}) and under-ice motion is less turbulent than the one with a free-surface, we may expect to have $A'_H$ values even smaller than $10\ m^2\, s^{-1}$ in this case.\\
Moreover, in the left-hand side of \eqref{NS corrected 1} we have also added the terms $2\boldsymbol{\Omega}'\times\bu'$ (Coriolis acceleration) and $\boldsymbol{\Omega}'\times(\boldsymbol{\Omega}'\times \boldsymbol{r}')$ (centrifugal acceleration, $\boldsymbol{r}'$ being the position vector) due to the Earth's rotation around its vertical axis $\mathbf{e}_3$ with angular with angular velocity $\Omega' \approx 7.29\cdot10^{-5} \, rad\,s^{-1}$, $\boldsymbol{\Omega}'=\Omega'\mathbf{e}_3$.\\
Equation \eqref{NS corrected 1} is consistent with the standard one in Cartesian coordinates \eqref{kunduNS}.
\subsubsection{The continuity equation in classical spherical coordinates}
The continuity equation $\frac{D\rho'}{D t'} +\rho' \nabla'\cdot \mathbf{u'}=0$ in spherical coordinates, using the above mentioned formula for the divergence, reads 
\begin{equation}
  \frac{D\rho'}{D t'} +\rho'\left[ \frac{1}{r' \cos \theta} \frac{\partial u'}{\partial \varphi} + \frac{1}{r' \cos \theta} \frac{\partial}{\partial \theta} \left( v' \cos \theta  \right) + \frac{1}{r'^2} \frac{\partial}{\partial r'} \left( r'^2 w' \right)\right]=0.
\end{equation}

 \subsection{Rotated Spherical Coordinates}\label{rotated spherical}

\begin{figure}
    \centering
    \includegraphics[width=0.5\linewidth]{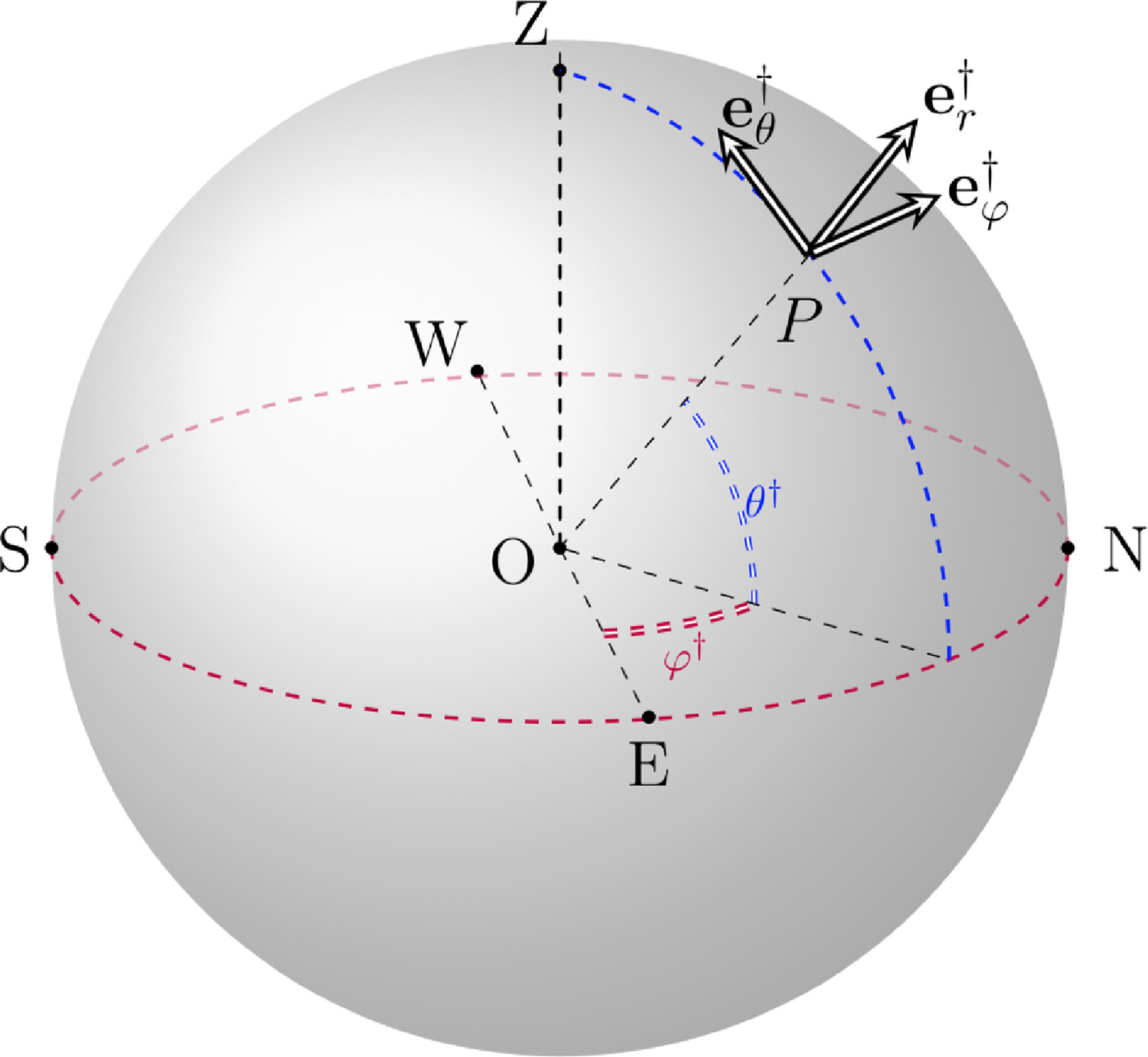}
    \caption{Rotated spherical coordinate system. Image from \cite{CJ2023}. CC BY 4.0 (\url{http://creativecommons.org/licenses/by/4.0})}
    \label{fig-rotated}
\end{figure}

Classical spherical coordinates present two polar singularities, where longitude is undefined at the North and South Poles. This issue reflects the fundamental impossibility of defining global coordinates on a sphere: singularities, where the coordinate system either fails or degenerates, are inevitable. According to the Poincaré–Hopf index theorem, continuous vector fields free of singularities can exist only on manifolds with a zero Euler characteristic. Since the sphere has an Euler characteristic of 2, there cannot exist a continuous, nonvanishing tangent vector field on it. This result is commonly referred to as the ``hairy ball theorem" (see \cite{17}).\\
The convergence of meridians at the poles illustrates this limitation in spherical coordinates. However, the problem can be resolved by rotating the coordinate system so that both singularities are relocated to the Equator. This adjustment enables a smooth parameterization for the entire region north (or south) of the Equator. In this section we review the construction of this rotated spherical coordinates developed in \cite{CJ2023}. See also \cite{CJChapter}.\\
Start by considering the standard Cartesian coordinate system $(\mathbf{e}_1, \mathbf{e}_2, \mathbf{e}_3) $ positioned at the center of the Earth, pointing in the direction $\Vec{OZ}$ of Null Island,  $\Vec{OE}$ East,  $\Vec{ON}$ North, respectively. See Figure \ref{fig-rotated}.\\
Now, define a new Cartesian coordinate system  $(\mathbf{e}_1^{\dag}, \mathbf{e}_2^{\dag}, \mathbf{e}_3^{\dag}) $ by permuting cyclically the first three Cartesian axes
\begin{equation}\label{4.3}
\mathbf{e}_1^{\dag}=\ec,\quad \mathbf{e}_2^{\dag}=\ea, \quad\mathbf{e}_3^{\dag}=\eb.
\end{equation}
In terms of the associated azimuthal $\theta^{\dag}\in[-\frac{\pi}{2},\frac{\pi}{2}]$ and meridional $\varphi^{\dag}\in[0,2\pi)$  angles, the coordinates of a point $P$ on the Earth are
\begin{equation}
r'\cos\td\cos\pd \ea^{\dag}+r'\cos\td\sin\pd \eb^{\dag}+r'\sin\td \ec^{\dag},
\end{equation}
while in the standard system they are
\begin{equation}
r'\cos\theta\cos\varphi \ea+r'\cos\theta\sin\varphi \eb+r'\sin\theta \ec.
\end{equation}
Therefore, using \eqref{4.3}, it follows that
\begin{equation}\label{rot identities}
\left\{\begin{aligned}
&\cos\theta^{\dag}\cos\varphi^{\dag}=\cos\theta\sin\varphi,\\
&\cos\theta^{\dag}\sin\varphi^{\dag}=\sin\theta,\\
&\sin\theta^{\dag}=\cos\theta\cos\varphi.
\end{aligned}\right.
\end{equation}
In this new coordinate system the North Pole has coordinates $\pd=\frac{\pi}{2},\ \td=0$, and the solution of the system \eqref{rot identities} for the Arctic Ocean is 
\begin{equation}
\left\{\begin{aligned}
&\pd=\cot^{-1}(\sin\varphi \cot\theta) \in (0,\pi),\\
&\td=\sin^{-1}(\cos\varphi \cos\theta) \in \left(-\frac{\pi}{2},\frac{\pi}{2}\right),
\end{aligned}\right.
\end{equation}
The unit vectors in the rotated spherical coordinates $(\pd, \td, r')$ are  $(\ep^{\dag}, \et^{\dag}, \er)$, given by

 \begin{equation}
\left\{ \begin{array}{ll}
  \ep^{\dag}=-\sin\varphi^{\dag} \ea^{\dag}+\cos\varphi^{\dag} \eb^{\dag},\\
\et^{\dag}=-\cos\varphi^{\dag}\sin\theta^{\dag} \ea^{\dag}-\sin\varphi^{\dag}\sin\theta^{\dag} \eb^{\dag}+\cos\theta^{\dag} \ec^{\dag},\\
    \er^{\dag}=\cos\varphi^{\dag}\cos\theta^{\dag} \ea^{\dag}+\sin\varphi^{\dag}\cos\theta^{\dag} \eb^{\dag}+\sin\theta^{\dag} \ec^{\dag},
 \end{array}\right.
  \label{base sferiche dag}
\end{equation}
and the corresponding velocity components are $(u'_{\dag}, v'_{\dag}, w'_{\dag})$.\\
In the new rotated spherical coordinates the momentum equations are
\begin{equation}\label{NS rotated}
\begin{aligned}
& \rho' \frac{D} {D t'}\begin{pmatrix}
u'_{\dag} \\
v'_{\dag}\\
w'_{\dag}\\
\end{pmatrix}+\frac{\rho'}{r'}\begin{pmatrix}-u'_{\dag}v'_{\dag}\tan \theta^{\dag} + u'_{\dag}w'_{\dag}\\
u'^2_{\dag}\tan\theta^{\dag}+v'_{\dag}w'_{\dag}\\
-u'^2_{\dag}-v'^2_{\dag}
\end{pmatrix}+\\
&\quad +2\rho'\Omega'\begin{pmatrix}
    -v'_{\dag}\sin\varphi^{\dag}\cos\theta^{\dag}-w'_{\dag}\sin\varphi^{\dag}\sin\theta^{\dag}\\
    u'_{\dag}\sin\varphi^{\dag}\cos\theta^{\dag}-w'_{\dag}\cos\varphi^{\dag}\\
   u'_{\dag} \sin\varphi^{\dag}\sin\theta^{\dag} +v'_{\dag} \cos\varphi^{\dag}
\end{pmatrix}+\\
&\quad +\rho' r' \Omega'^2\begin{pmatrix}
    \sin\varphi^{\dag}\cos\varphi^{\dag}\cos\theta^{\dag}\\
    -\sin^2\varphi^{\dag} \sin\theta^{\dag}\cos\theta^{\dag}\\
   -\cos^2\varphi^{\dag}\cos^2\theta^{\dag} -\sin^2\theta^{\dag}
\end{pmatrix} = \\
& =-\nabla'p' +\rho\begin{pmatrix}
    0\\
    0\\
     -g'\frac{R'^2}{ r'^2}\\
\end{pmatrix}
+\Delta'_{\mu}
\begin{pmatrix}
    u'_{\dag}\\ v'_{\dag}\\ w'_{\dag}
\end{pmatrix}-\frac{1}{3}\begin{pmatrix}
    \frac{1}{r'\cos \theta^{\dag}}\frac{\partial}{\partial \varphi^{\dag}}\left(\frac{\mu'_H}{\rho'} \frac{D\rho'}{D t'}\right)\\
    \frac{1}{r'}\frac{\partial}{\partial \theta^{\dag}}\left(\frac{\mu'_H}{\rho'} \frac{D\rho'}{D t'}\right)\\
    \frac{\partial}{\partial r'}\left(\frac{\mu'_V}{\rho'} \frac{D\rho'}{D t'}\right)
\end{pmatrix}-\\
& \quad-\frac{1}{r'^2\cos^2\theta^{\dag}}\begin{pmatrix}
    \mu'_H u'_{\dag}\\
    \mu'_H v'_{\dag}\\
    2\mu'_V(w'_{\dag}\cos^2\theta^{\dag}-v'_{\dag}\sin\theta^{\dag}\cos\theta^{\dag})
\end{pmatrix}+\\
& \quad+ \frac{2\mu'_H}{r'^2}\frac{\partial}{\partial \theta^{\dag}}\begin{pmatrix}
0\\
 w'_{\dag}\\
   -v'_{\dag}
\end{pmatrix}+\frac{2\mu'_H}{r'^2\cos\theta^{\dag}}\frac{\partial}{\partial \varphi^{\dag}}\begin{pmatrix}
w'_{\dag}-v'_{\dag}\tan\theta^{\dag}\\
 u'_{\dag}\tan\theta^{\dag}\\
   -u'_{\dag}
\end{pmatrix}+ \\
&\quad +\frac{d \mu'_V}{dr'}r' \begin{pmatrix}
    \frac{\partial}{\partial r'}\left(\frac{u'_{\dag}}{r'}\right)\\
       \frac{\partial}{\partial r'}\left(\frac{v'_{\dag}}{r'}\right)\\
        {0}
\end{pmatrix}+\frac{d \mu'_H}{d r'} \begin{pmatrix}
    \frac{1}{r'\cos \theta}\frac{\partial  {w'_{\dag}}}{\partial \varphi^{\dag}}\\
     \frac{1}{r'} \frac{\partial {w'_{\dag}}}{\partial \theta^{\dag}}\\
{0}
\end{pmatrix}+ \frac{d \mu'_V}{dr'} \begin{pmatrix}
   0\\
   0\\
   2 \frac{\partial w'_{\dag}}{\partial r'}  
\end{pmatrix}+\\
&\quad +\frac{d \mu'_V}{dr'}\begin{pmatrix}
   0\\
   0\\
   \frac{1}{r' \cos \theta^{\dag}} \frac{\partial u'_{\dag}}{\partial \varphi^{\dag}} + \frac{1}{r' \cos \theta^{\dag}} \frac{\partial}{\partial \theta^{\dag}} \left( v'_{\dag} \cos \theta^{\dag}  \right) + \frac{1}{r'^2} \frac{\partial}{\partial r'} \left( r'^2 w'_{\dag} \right)
\end{pmatrix},
\end{aligned}
\end{equation}
where 
\begin{equation}
	\begin{aligned}
		\nabla'&= \left( \frac{1}{r'\cos\theta^{\dag}}\frac{\partial}{\partial\varphi^{\dag}},\  \frac{1}{r'}\frac{\partial}{\partial\theta^{\dag}} ,\   \frac{\partial}{\partial r'}
		\right),\\
		\frac{D}{Dt'}&=\frac{\partial} {\partial t'}+ \frac{u'_{\dag}}{r' \cos \theta^{\dag}} \frac{\partial }{\partial \varphi^{\dag}} + \frac{v'_{\dag}}{r'} \frac{\partial }{\partial \theta^{\dag}} + w'_{\dag}\frac{\partial }{\partial r'}, \\
		\Delta'_{\mu}&=\mu'_V \left(\frac{\partial^2}{\partial r'^2}+ \frac{2}{r'}\frac{\partial}{\partial r'}\right) + \frac{\mu'_H}{r'^2}\left(\frac{1}{ \cos^2 \theta^{\dag}} \frac{\partial^2 }{\partial \varphi^{\dag 2}}+\frac{\partial^2}{\partial \theta^{\dag 2}}   -  \tan\theta^{\dag}\frac{\partial}{\partial \theta^{\dag}}\right),
	\end{aligned}
\end{equation}
while the continuity equation is
\begin{equation}\label{mass rot}
\frac{D'\rho}{Dt'}+\rho'\left[\frac{1}{r' \cos \theta^{\dag}} \frac{\partial u'_{\dag}}{\partial \varphi^{\dag}} + \frac{1}{r' \cos \theta^{\dag}} \frac{\partial}{\partial \theta^{\dag}} \left( v'_{\dag} \cos \theta^{\dag}  \right) + \frac{1}{r'^2} \frac{\partial}{\partial r'} \left( r'^2 w'_{\dag}\right)\right]=0.
%
\end{equation}
The Navier-Stokes equations in the new spherical coordinate system \eqref{NS rotated} look very similar to the ones in the classical spherical system \eqref{NS corrected 1}, except for two points: in \eqref{NS rotated} the Coriolis acceleration and the centrifugal acceleration terms are now given by, respectively,
\begin{equation}
	\begin{aligned}	
&2\mathbf{\Omega}'\times\mathbf{u}'=2\Omega'\begin{pmatrix}
    -v'_{\dag}\sin\varphi^{\dag}\cos\theta^{\dag}-w'_{\dag}\sin\varphi^{\dag}\sin\theta^{\dag}\\
    u'_{\dag}\sin\varphi^{\dag}\cos\theta^{\dag}-w'_{\dag}\cos\varphi^{\dag}\\
   u'_{\dag} \sin\varphi^{\dag}\sin\theta^{\dag} +v'_{\dag} \cos\varphi^{\dag}
\end{pmatrix},\\ 
&\mathbf{\Omega}'\times(\mathbf{\Omega}'\times \mathbf{r}')= r' \Omega'^2\begin{pmatrix}
    \sin\varphi^{\dag}\cos\varphi^{\dag}\cos\theta^{\dag}\\
    -\sin^2\varphi^{\dag} \sin\theta^{\dag}\cos\theta^{\dag}\\
   -\cos^2\varphi^{\dag}\cos^2\theta^{\dag} -\sin^2\theta^{\dag}
\end{pmatrix},
\end{aligned}
\end{equation}
but more importantly, we remark that \eqref{NS corrected 1} does not make sense at the North (and South) Pole, while \eqref{NS rotated} does (and vice-versa \eqref{NS corrected 1} is correct at the Equator, while\eqref{NS rotated} is not defined there).\\
Finally, although we have not make use of this in the analysis of our interest, we remark that, since $\mathbf{\Omega}'\times(\mathbf{\Omega}'\times \mathbf{r}')=\frac{1}{2}\nabla'(\Omega'^2\ell'^2)=:\nabla'\Phi'$, it is common to  write the pressure as
\begin{equation}
	P'=p'-\rho' g' \frac{R'^2 }{r'}+\rho'\Phi',
\end{equation}
where $\ell'$ is the distance of the point from the axis or rotation $\ec$, and $\Phi'=-\frac{1}{2} r'^2\Omega'^2 (\cos^2\varphi^{\dag}\cos^2\theta^{\dag} + \sin^2\theta^{\dag})$.\\
The momentum equations with the re-defined pressure therefore become
\begin{equation}\label{NS rotated 2}
	\begin{aligned}
		& \rho' \frac{D} {D t'}\begin{pmatrix}
			u'_{\dag} \\
			v'_{\dag}\\
			w'_{\dag}\\
		\end{pmatrix}+\frac{\rho'}{r'}\begin{pmatrix}-u'_{\dag}v'_{\dag}\tan \theta^{\dag} + u'_{\dag}w'_{\dag}\\
			u'^2_{\dag}\tan\theta^{\dag}+v'_{\dag}w'_{\dag}\\
			-u'^2_{\dag}-v'^2_{\dag}
		\end{pmatrix}+\\
        &\quad + 2\rho'\Omega'\begin{pmatrix}
			-v'_{\dag}\sin\varphi^{\dag}\cos\theta^{\dag}-w'_{\dag}\sin\varphi^{\dag}\sin\theta^{\dag}\\
			u'_{\dag}\sin\varphi^{\dag}\cos\theta^{\dag}-w'_{\dag}\cos\varphi^{\dag}\\
			u'_{\dag} \sin\varphi^{\dag}\sin\theta^{\dag} +v'_{\dag} \cos\varphi^{\dag}
		\end{pmatrix}=\\
		& =-\nabla'P'-\nabla'\rho'\left(\Phi'+g'\frac{R'^2}{ r'}\right) 
		+\Delta'_{\mu}
		\begin{pmatrix}
			u'_{\dag}\\ v'_{\dag}\\ w'_{\dag}
		\end{pmatrix}-\frac{1}{3}\begin{pmatrix}
			\frac{1}{r'\cos \theta^{\dag}}\frac{\partial}{\partial \varphi^{\dag}}\left(\frac{\mu'_H}{\rho'} \frac{D\rho'}{D t'}\right)\\
			\frac{1}{r'}\frac{\partial}{\partial \theta^{\dag}}\left(\frac{\mu'_H}{\rho'} \frac{D\rho'}{D t'}\right)\\
			\frac{\partial}{\partial r'}\left(\frac{\mu'_V}{\rho'} \frac{D\rho'}{D t'}\right)
		\end{pmatrix}-\\
		&\quad -\frac{1}{r'^2\cos^2\theta^{\dag}}\begin{pmatrix}
			\mu'_H u'_{\dag}\\
			\mu'_H v'_{\dag}\\
			2\mu'_V(w'_{\dag}\cos^2\theta^{\dag}-v'_{\dag}\sin\theta^{\dag}\cos\theta^{\dag})
		\end{pmatrix}+\\
		& \quad+ \frac{2\mu'_H}{r'^2}\frac{\partial}{\partial \theta^{\dag}}\begin{pmatrix}
			0\\
			w'_{\dag}\\
			-v'_{\dag}
		\end{pmatrix}+\frac{2\mu'_H}{r'^2\cos\theta^{\dag}}\frac{\partial}{\partial \varphi^{\dag}}\begin{pmatrix}
			w'_{\dag}-v'_{\dag}\tan\theta^{\dag}\\
			u'_{\dag}\tan\theta^{\dag}\\
			-u'_{\dag}
		\end{pmatrix}+\\
        &\quad+\frac{d \mu'_V}{dr'}r' \begin{pmatrix}
			\frac{\partial}{\partial r'}\left(\frac{u'_{\dag}}{r'}\right)\\
			\frac{\partial}{\partial r'}\left(\frac{v'_{\dag}}{r'}\right)\\
			{0}
		\end{pmatrix}+\frac{d \mu'_H}{d r'} \begin{pmatrix}
			\frac{1}{r'\cos \theta}\frac{\partial  {w'_{\dag}}}{\partial \varphi^{\dag}}\\
			\frac{1}{r'} \frac{\partial {w'_{\dag}}}{\partial \theta^{\dag}}\\
			{0}
		\end{pmatrix}+ \frac{d \mu'_V}{dr'} \begin{pmatrix}
			0\\
			0\\
			2 \frac{\partial w'_{\dag}}{\partial r'}  
		\end{pmatrix}+\\
        &\quad+\frac{d \mu'_V}{dr'}\begin{pmatrix}
			0\\
			0\\
			\frac{1}{r' \cos \theta^{\dag}} \frac{\partial u'_{\dag}}{\partial \varphi^{\dag}} + \frac{1}{r' \cos \theta^{\dag}} \frac{\partial}{\partial \theta^{\dag}} \left( v'_{\dag} \cos \theta^{\dag}  \right) + \frac{1}{r'^2} \frac{\partial}{\partial r'} \left( r'^2 w'_{\dag} \right) 
		\end{pmatrix}.
	\end{aligned}
\end{equation}
This approach is particularly useful when considering constant density, in which case the term $\nabla'\rho'\left(\Phi'+g'\frac{R'^2}{ r'}\right)$ is zero.

\end{appendices}

\bibliography{sn-bibliography}


\begin{thebibliography}{36}
\ifx \bisbn   \undefined \def \bisbn  #1{ISBN #1}\fi
\ifx \binits  \undefined \def \binits#1{#1}\fi
\ifx \bauthor  \undefined \def \bauthor#1{#1}\fi
\ifx \batitle  \undefined \def \batitle#1{#1}\fi
\ifx \bjtitle  \undefined \def \bjtitle#1{#1}\fi
\ifx \bvolume  \undefined \def \bvolume#1{\textbf{#1}}\fi
\ifx \byear  \undefined \def \byear#1{#1}\fi
\ifx \bissue  \undefined \def \bissue#1{#1}\fi
\ifx \bfpage  \undefined \def \bfpage#1{#1}\fi
\ifx \blpage  \undefined \def \blpage #1{#1}\fi
\ifx \burl  \undefined \def \burl#1{\textsf{#1}}\fi
\ifx \doiurl  \undefined \def \doiurl#1{\url{https://doi.org/#1}}\fi
\ifx \betal  \undefined \def \betal{\textit{et al.}}\fi
\ifx \binstitute  \undefined \def \binstitute#1{#1}\fi
\ifx \binstitutionaled  \undefined \def \binstitutionaled#1{#1}\fi
\ifx \bctitle  \undefined \def \bctitle#1{#1}\fi
\ifx \beditor  \undefined \def \beditor#1{#1}\fi
\ifx \bpublisher  \undefined \def \bpublisher#1{#1}\fi
\ifx \bbtitle  \undefined \def \bbtitle#1{#1}\fi
\ifx \bedition  \undefined \def \bedition#1{#1}\fi
\ifx \bseriesno  \undefined \def \bseriesno#1{#1}\fi
\ifx \blocation  \undefined \def \blocation#1{#1}\fi
\ifx \bsertitle  \undefined \def \bsertitle#1{#1}\fi
\ifx \bsnm \undefined \def \bsnm#1{#1}\fi
\ifx \bsuffix \undefined \def \bsuffix#1{#1}\fi
\ifx \bparticle \undefined \def \bparticle#1{#1}\fi
\ifx \barticle \undefined \def \barticle#1{#1}\fi
\bibcommenthead
\ifx \bconfdate \undefined \def \bconfdate #1{#1}\fi
\ifx \botherref \undefined \def \botherref #1{#1}\fi
\ifx \url \undefined \def \url#1{\textsf{#1}}\fi
\ifx \bchapter \undefined \def \bchapter#1{#1}\fi
\ifx \bbook \undefined \def \bbook#1{#1}\fi
\ifx \bcomment \undefined \def \bcomment#1{#1}\fi
\ifx \oauthor \undefined \def \oauthor#1{#1}\fi
\ifx \citeauthoryear \undefined \def \citeauthoryear#1{#1}\fi
\ifx \endbibitem  \undefined \def \endbibitem {}\fi
\ifx \bconflocation  \undefined \def \bconflocation#1{#1}\fi
\ifx \arxivurl  \undefined \def \arxivurl#1{\textsf{#1}}\fi
\csname PreBibitemsHook\endcsname

\bibitem[\protect\citeauthoryear{Jakobsson et~al.}{2012}]{IBCAO}
\begin{barticle}
\bauthor{\bsnm{Jakobsson}, \binits{M.}}, \betal:
\batitle{The {I}nternational {B}athymetric {C}hart of the {A}rctic {O}cean
  ({IBCAO}) version 3.0}.
\bjtitle{Geophys. Res. Lett.}
(\byear{2012})
\doiurl{10.1029/2012GL052219}
\end{barticle}
\endbibitem

\bibitem[\protect\citeauthoryear{Timmermans and Marshall}{2020}]{TM2020}
\begin{barticle}
\bauthor{\bsnm{Timmermans}, \binits{M.-L.}},
\bauthor{\bsnm{Marshall}, \binits{J.}}:
\batitle{Understanding {A}rctic {O}cean circulation: a review of ocean dynamics
  in a changing climate}.
\bjtitle{J. Geophys. Res. Oceans}
\bvolume{125},
\bfpage{2018}--\blpage{014378}
(\byear{2020})
\doiurl{10.1029/2018JC014378}
\end{barticle}
\endbibitem

\bibitem[\protect\citeauthoryear{Johnson}{2022}]{Johnson2022}
\begin{barticle}
\bauthor{\bsnm{Johnson}, \binits{R.S.}}:
\batitle{The ocean and the atmosphere: {A}n applied mathematician's view}.
\bjtitle{Commun. Pure Appl. Math.}
\bvolume{21}(\bissue{7}),
\bfpage{2357}--\blpage{2381}
(\byear{2022})
\doiurl{10.3934/cpaa.2022040}
\end{barticle}
\endbibitem

\bibitem[\protect\citeauthoryear{Constantin and Johnson}{2023}]{CJ2023}
\begin{barticle}
\bauthor{\bsnm{Constantin}, \binits{A.}},
\bauthor{\bsnm{Johnson}, \binits{R.S.}}:
\batitle{On the dynamics of the near-surface currents in the {A}rctic {O}cean}.
\bjtitle{Nonlinear Anal., Real World Appl.}
\bvolume{73},
\bfpage{103894}
(\byear{2023})
\doiurl{10.1016/j.nonrwa.2023.103894}
\end{barticle}
\endbibitem

\bibitem[\protect\citeauthoryear{Constantin and Johnson}{2024}]{CJ2024}
\begin{barticle}
\bauthor{\bsnm{Constantin}, \binits{A.}},
\bauthor{\bsnm{Johnson}, \binits{R.S.}}:
\batitle{The dynamics of the transpolar drift current}.
\bjtitle{Geophys. Astrophys. Fluid Dyn.}
\bvolume{118}(\bissue{3}),
\bfpage{165}--\blpage{182}
(\byear{2024})
\doiurl{10.1080/03091929.2024.2351919}
\end{barticle}
\endbibitem

\bibitem[\protect\citeauthoryear{Johnson}{2024}]{Johnson2024}
\begin{barticle}
\bauthor{\bsnm{Johnson}, \binits{R.S.}}:
\batitle{The {T}ranspolar {D}rift current: an ocean-ice-wind complex in
  rotating, spherical coordinates}.
\bjtitle{Monatsh. Math.}
\bvolume{205}(\bissue{4}),
\bfpage{735}--\blpage{755}
(\byear{2024})
\doiurl{10.1007/s00605-024-01995-7}
\end{barticle}
\endbibitem

\bibitem[\protect\citeauthoryear{Constantin}{2022a}]{Constantin2022}
\begin{barticle}
\bauthor{\bsnm{Constantin}, \binits{A.}}:
\batitle{Nonlinear wind-drift ocean currents in arctic regions}.
\bjtitle{Geophys. Astrophys. Fluid Dyn.}
\bvolume{116}(\bissue{2}),
\bfpage{101}--\blpage{115}
(\byear{2022})
\doiurl{10.1080/03091929.2021.1981307}
\end{barticle}
\endbibitem

\bibitem[\protect\citeauthoryear{Constantin}{2022b}]{Constantin2022NOTE}
\begin{barticle}
\bauthor{\bsnm{Constantin}, \binits{A.}}:
\batitle{Comments on: nonlinear wind-drift ocean currents in arctic regions}.
\bjtitle{Geophys. Astrophys. Fluid Dyn.}
\bvolume{116}(\bissue{2}),
\bfpage{116}--\blpage{121}
(\byear{2022})
\doiurl{10.1080/03091929.2022.2036337}
\end{barticle}
\endbibitem

\bibitem[\protect\citeauthoryear{Constantin and Johnson}{2024}]{CJChapter}
\begin{bchapter}
\bauthor{\bsnm{Constantin}, \binits{A.}},
\bauthor{\bsnm{Johnson}, \binits{R.S.}}:
\bctitle{Spherical coordinates for {A}rctic {O}cean flows}.
In: \beditor{\bsnm{Henry}, \binits{D.}} (ed.)
\bbtitle{Nonlinear Dispersive Waves, Advances in Mathematical Fluid Mechanics},
pp. \bfpage{239}--\blpage{282}.
\bpublisher{Springer},
\blocation{Cham}
(\byear{2024}).
\doiurl{10.1007/978-3-031-63512-0_11}
\end{bchapter}
\endbibitem

\bibitem[\protect\citeauthoryear{Ekman}{1905}]{Ekman1905}
\begin{barticle}
\bauthor{\bsnm{Ekman}, \binits{V.W.}}:
\batitle{On the influence of the {E}arth's rotation on ocean currents}.
\bjtitle{Arkiv. Mat. Astron. Fys.}
\bvolume{2},
\bfpage{1}--\blpage{52}
(\byear{1905})
\end{barticle}
\endbibitem

\bibitem[\protect\citeauthoryear{Jenkins and Bye}{2006}]{JenkinsB2006}
\begin{barticle}
\bauthor{\bsnm{Jenkins}, \binits{A.D.}},
\bauthor{\bsnm{Bye}, \binits{J.A.T.}}:
\batitle{Some aspects of the work of {V.W. E}kman}.
\bjtitle{Polar Record}
\bvolume{42}(\bissue{1}),
\bfpage{15}--\blpage{22}
(\byear{2006})
\doiurl{10.1017/S0032247405004845}
\end{barticle}
\endbibitem

\bibitem[\protect\citeauthoryear{Constantin}{2020}]{Constantin2020}
\begin{barticle}
\bauthor{\bsnm{Constantin}, \binits{A.}}:
\batitle{Frictional effects in wind-driven ocean currents}.
\bjtitle{Geophys. Astrophys. Fluid Dyn.}
\bvolume{115}(\bissue{1}),
\bfpage{1}--\blpage{14}
(\byear{2020})
\doiurl{10.1080/03091929.2020.1748614}
\end{barticle}
\endbibitem

\bibitem[\protect\citeauthoryear{Roberti}{2022}]{Roberti}
\begin{barticle}
\bauthor{\bsnm{Roberti}, \binits{L.}}:
\batitle{The {E}kman spiral for piecewise-constant eddy viscosity}.
\bjtitle{Appl. Anal.}
\bvolume{101}(\bissue{15}),
\bfpage{5528}--\blpage{5536}
(\byear{2022})
\doiurl{10.1080/00036811.2021.1896709}
\end{barticle}
\endbibitem

\bibitem[\protect\citeauthoryear{Shrira and Almelah}{2020}]{ShriraJFM}
\begin{barticle}
\bauthor{\bsnm{Shrira}, \binits{V.I.}},
\bauthor{\bsnm{Almelah}, \binits{R.B.}}:
\batitle{Upper-ocean {E}kman current dynamics: a new perspective}.
\bjtitle{J. Fluid Mech.}
\bvolume{887},
\bfpage{24}
(\byear{2020})
\doiurl{10.1017/jfm.2019.1059}
\end{barticle}
\endbibitem

\bibitem[\protect\citeauthoryear{Soloviev and Lukas}{2014}]{Soloviev2006}
\begin{bbook}
\bauthor{\bsnm{Soloviev}, \binits{A.}},
\bauthor{\bsnm{Lukas}, \binits{R.}}:
\bbtitle{The {N}ear-{S}urface {L}ayer of the {O}cean. {S}tructure, {D}ynamics
  And {A}pplications}.
\bpublisher{Springer},
\blocation{Dordrecht}
(\byear{2014})
\end{bbook}
\endbibitem

\bibitem[\protect\citeauthoryear{Constantin and Johnson}{2021}]{CJ2021}
\begin{barticle}
\bauthor{\bsnm{Constantin}, \binits{A.}},
\bauthor{\bsnm{Johnson}, \binits{R.S.}}:
\batitle{On the modelling of large-scale atmospheric flows}.
\bjtitle{J. Diff. Eq.}
\bvolume{285},
\bfpage{751}--\blpage{79}
(\byear{2021})
\doiurl{10.1016/j.jde.2021.03.019}
\end{barticle}
\endbibitem

\bibitem[\protect\citeauthoryear{Roquet et~al.}{2015}]{Roquet}
\begin{barticle}
\bauthor{\bsnm{Roquet}, \binits{F.}},
\bauthor{\bsnm{Madec}, \binits{G.}},
\bauthor{\bsnm{Brodeau}, \binits{L.}},
\bauthor{\bsnm{Nycander}, \binits{J.}}:
\batitle{Defining a simplified yet “realistic” equation of state for
  seawater}.
\bjtitle{J. Phys. Oceanogr.}
\bvolume{45},
\bfpage{2564}--\blpage{2579}
(\byear{2015})
\doiurl{10.1175/JPO-D-15-0080.1}
\end{barticle}
\endbibitem

\bibitem[\protect\citeauthoryear{Talley et~al.}{2011}]{Talley}
\begin{bbook}
\bauthor{\bsnm{Talley}, \binits{L.D.}},
\bauthor{\bsnm{Pickard}, \binits{G.L.}},
\bauthor{\bsnm{Emery}, \binits{W.J.}},
\bauthor{\bsnm{Swift}, \binits{J.H.}}:
\bbtitle{Descriptive {P}hysical {O}ceanography: {A}n {I}ntroduction}.
\bpublisher{Elsevier},
\blocation{Amsterdam}
(\byear{2011})
\end{bbook}
\endbibitem

\bibitem[\protect\citeauthoryear{Cavallini and Crisciani}{2013}]{CavalliniC}
\begin{bbook}
\bauthor{\bsnm{Cavallini}, \binits{F.}},
\bauthor{\bsnm{Crisciani}, \binits{F.}}:
\bbtitle{Quasi-geostrophic Theory of Oceans and Atmosphere: Topics in the
  Dynamics and Thermodynamics of the Fluid Earth}.
\bpublisher{Springer},
\blocation{Dordrecht}
(\byear{2013})
\end{bbook}
\endbibitem

\bibitem[\protect\citeauthoryear{Meincke et~al.}{1997}]{Meincke}
\begin{barticle}
\bauthor{\bsnm{Meincke}, \binits{J.}},
\bauthor{\bsnm{Rudels}, \binits{B.}},
\bauthor{\bsnm{Friedrich}, \binits{H.J.}}:
\batitle{The {A}rctic {O}cean-{N}ordic {S}eas thermohaline system}.
\bjtitle{J. Mar. Sci.}
\bvolume{54}(\bissue{3}),
\bfpage{283}--\blpage{299}
(\byear{1997})
\doiurl{10.1006/jmsc.1997.0229}
\end{barticle}
\endbibitem

\bibitem[\protect\citeauthoryear{Rawlins et~al.}{2010}]{Rawlins}
\begin{barticle}
\bauthor{\bsnm{Rawlins}, \binits{M.A.}}, \betal:
\batitle{Analysis of the {A}rctic system for freshwater cycle intensiﬁcation:
  {O}bservations and expectations}.
\bjtitle{J. Clim.}
\bvolume{23}(\bissue{21}),
\bfpage{5715}--\blpage{5737}
(\byear{2010})
\doiurl{10.1175/2010JCLI3421.1}
\end{barticle}
\endbibitem

\bibitem[\protect\citeauthoryear{Overeem and Syvitski}{2010}]{Overeem}
\begin{barticle}
\bauthor{\bsnm{Overeem}, \binits{I.}},
\bauthor{\bsnm{Syvitski}, \binits{J.P.M.}}:
\batitle{Shifting discharge peaks in {A}rctic rivers, 1977–2007}.
\bjtitle{Geogr. Ann., Ser. A}
\bvolume{92}(\bissue{2}),
\bfpage{285}--\blpage{296}
(\byear{2010})
\doiurl{10.1111/j.1468-0459.2010.00395.x}
\end{barticle}
\endbibitem

\bibitem[\protect\citeauthoryear{Wadhams}{2002}]{Wadhams}
\begin{bbook}
\bauthor{\bsnm{Wadhams}, \binits{P.}}:
\bbtitle{Ice in the {O}cean}.
\bpublisher{CRC Press},
\blocation{Boca Raton}
(\byear{2002})
\end{bbook}
\endbibitem

\bibitem[\protect\citeauthoryear{Yang}{2006}]{Yang}
\begin{barticle}
\bauthor{\bsnm{Yang}, \binits{J.}}:
\batitle{The seasonal variability of the {A}rctic {O}cean {E}kman transport and
  its role in the mixed layer heat and salt fluxes}.
\bjtitle{J. Clim.}
\bvolume{19},
\bfpage{5366}--\blpage{5387}
(\byear{2006})
\doiurl{10.1175/JCLI3892.1}
\end{barticle}
\endbibitem

\bibitem[\protect\citeauthoryear{Kundu}{1990}]{Kundu}
\begin{bbook}
\bauthor{\bsnm{Kundu}, \binits{P.K.}}:
\bbtitle{Fluid {M}echanics},
\bedition{1}st edn.
\bpublisher{Academic Press},
\blocation{San Diego}
(\byear{1990})
\end{bbook}
\endbibitem

\bibitem[\protect\citeauthoryear{Abrashkin and
  Pelinovsky}{2022}]{AbrashkinPelinovsky}
\begin{barticle}
\bauthor{\bsnm{Abrashkin}, \binits{A.}},
\bauthor{\bsnm{Pelinovsky}, \binits{E.N.}}:
\batitle{Gerstner waves and their generalizations in hydrodynamics and
  geophysics}.
\bjtitle{Phys. Usp.}
\bvolume{65},
\bfpage{453}--\blpage{467}
(\byear{2022})
\doiurl{10.3367/UFNe.2021.05.038980}
\end{barticle}
\endbibitem

\bibitem[\protect\citeauthoryear{Milne-Thomson}{1938}]{Milne}
\begin{bbook}
\bauthor{\bsnm{Milne-Thomson}, \binits{L.M.}}:
\bbtitle{Theoretical {H}ydrodynamics},
\bedition{1}st edn.
\bpublisher{The Macmillan Co.},
\blocation{London}
(\byear{1938})
\end{bbook}
\endbibitem

\bibitem[\protect\citeauthoryear{Pedlosky}{1987}]{Pedlosky}
\begin{bbook}
\bauthor{\bsnm{Pedlosky}, \binits{J.}}:
\bbtitle{Geophysical Fluid Dynamics},
\bedition{2}nd edn.
\bpublisher{Springer},
\blocation{New York}
(\byear{1987})
\end{bbook}
\endbibitem

\bibitem[\protect\citeauthoryear{Arfken et~al.}{1985}]{Arfken}
\begin{bbook}
\bauthor{\bsnm{Arfken}, \binits{G.B.}},
\bauthor{\bsnm{Weber}, \binits{H.J.}},
\bauthor{\bsnm{Harris}, \binits{F.E.}}:
\bbtitle{Mathematical {M}ethods for {P}hysicists. {A} Comprehensive guide},
\bedition{3}rd edn.
\bpublisher{Academic Press},
\blocation{San Diego}
(\byear{1985})
\end{bbook}
\endbibitem

\bibitem[\protect\citeauthoryear{Jog}{2015}]{JOG}
\begin{bbook}
\bauthor{\bsnm{Jog}, \binits{C.S.}}:
\bbtitle{Continuum {M}echanics: {F}oundations and {A}pplications of
  {M}echanics},
\bedition{3}rd edn.
\bpublisher{Cambridge University Press},
\blocation{Cambridge}
(\byear{2015})
\end{bbook}
\endbibitem

\bibitem[\protect\citeauthoryear{Spiegel et~al.}{2009}]{Spiegel}
\begin{bbook}
\bauthor{\bsnm{Spiegel}, \binits{M.}},
\bauthor{\bsnm{Lipschutz}, \binits{S.}},
\bauthor{\bsnm{Spellman}, \binits{D.}}:
\bbtitle{Vector {A}nalysis and an {I}ntroduction to {T}ensor {A}nalysis},
\bedition{2}nd edn.
\bsertitle{Schaum’s outline series}.
\bpublisher{Mc-Graw-Hill},
\blocation{New York}
(\byear{2009})
\end{bbook}
\endbibitem

\bibitem[\protect\citeauthoryear{Godinho and Nat\'ario}{2014}]{RiemannianGeom}
\begin{bbook}
\bauthor{\bsnm{Godinho}, \binits{L.}},
\bauthor{\bsnm{Nat\'ario}, \binits{J.}}:
\bbtitle{An Introduction to {R}iemannian {G}eometry. {W}ith {A}pplications to
  {M}echanics And {R}elativity}.
\bpublisher{Springer},
\blocation{Cham}
(\byear{2014}).
\doiurl{10.1007/978-3-319-08666-8}
\end{bbook}
\endbibitem

\bibitem[\protect\citeauthoryear{Holton and Hakim}{2013}]{DynamicMeteo}
\begin{bbook}
\bauthor{\bsnm{Holton}, \binits{J.R.}},
\bauthor{\bsnm{Hakim}, \binits{G.J.}}:
\bbtitle{An {I}ntroduction to {D}ynamic {M}eteorology},
\bedition{5}th edn.
\bpublisher{Academic Press},
\blocation{Cambridge}
(\byear{2013}).
\doiurl{10.1016/C2009-0-63394-8}
\end{bbook}
\endbibitem

\bibitem[\protect\citeauthoryear{Jiang et~al.}{2020}]{Jiang2020}
\begin{barticle}
\bauthor{\bsnm{Jiang}, \binits{N.}},
\bauthor{\bsnm{Layton}, \binits{W.}},
\bauthor{\bsnm{McLaughlin}, \binits{M.}},
\bauthor{\bsnm{Rong}, \binits{Y.}},
\bauthor{\bsnm{Zhao}, \binits{H.}}:
\batitle{On the {F}oundations of {E}ddy {V}iscosity {M}odels of {T}urbulence}.
\bjtitle{Fluids}
\bvolume{5}(\bissue{4}),
\bfpage{167}
(\byear{2020})
\doiurl{10.3390/fluids5040167}
\end{barticle}
\endbibitem

\bibitem[\protect\citeauthoryear{Darrigol}{2005}]{16}
\begin{bbook}
\bauthor{\bsnm{Darrigol}, \binits{O.}}:
\bbtitle{Worlds of {F}low}.
\bpublisher{Oxford University Press},
\blocation{Oxford}
(\byear{2005})
\end{bbook}
\endbibitem

\bibitem[\protect\citeauthoryear{Burns and Gidea}{2005}]{17}
\begin{bbook}
\bauthor{\bsnm{Burns}, \binits{K.}},
\bauthor{\bsnm{Gidea}, \binits{M.}}:
\bbtitle{Differential {G}eometry and {T}opology. {W}ith a {V}iew to {D}ynamical
  {S}ystems}.
\bpublisher{Chapman \& Hall/CRC},
\blocation{Boca Raton}
(\byear{2005})
\end{bbook}
\endbibitem

\end{thebibliography}

\end{document}